\documentclass[letterpaper,fleqn,usenatbib,usedcolumn]{mnras}

\usepackage[T1]{fontenc}
\usepackage{ae,aecompl}

\usepackage{graphicx}	
\usepackage{amsmath}	
\usepackage{amssymb}	
\usepackage{times}
\usepackage{breakurl}

\def\Halpha{{\rm H}\alpha}
\def\Hbeta{{\rm H}\beta}

\title[White dwarfs in the Galactic plane]{A search for white dwarfs in the Galactic plane: the field and the open cluster population }
\author[R. Raddi et al.]{R. Raddi$^{1}$\thanks{E-mail: r.raddi@warwick.ac.uk}, S. Catal\'{a}n$^{1}$, B. T. G\"ansicke$^{1}$, J.J. Hermes$^{1}$, R. Napiwotzki$^2$, D. Koester$^{3}$
\newauthor P.-E. Tremblay$^{1}$, G. Barentsen$^2$, H. J. Farnhill$^2$, M. Mohr-Smith$^2$,  J. E. Drew$^2$, 
\newauthor P. J. Groot$^{4}$, L. Guzman-Ramirez$^5$, Q. A. Parker$^{6,7,8}$, D. Steeghs$^{1}$, A. Zijlstra$^9$\\
$^{1}$Department of Physics, University of Warwick, Gibbet Hill Road, Coventry CV4 7AL, UK\\
$^{2}$Centre for Astrophysics Research, Science and Technology Research Institute, University of Hertfordshire, Hatfield AL10 9AB, UK \\
$^{3}$Institut f\"ur Theoretische Physik und Astrophysik, Universit\"at Kiel, 24098, Kiel, Germany\\
$^{4}$Department of Astrophysics, IMAPP, Radboud University Nijmegen, PO Box 9010, 6500 GL Nijmegen, the Netherlands\\
$^{5}$European Southern Observatory, Alonso de Cordova 3107, Vitacura, Santiago, Chile\\
$^{6}$Department of Physics, Chong Yeut Ming Physics building, The University of Hong Kong, Pokfulam, Hong Kong\\
$^{7}$Department of Physics \& Astronomy, Macquarie University, Sydney, NSW 2109, Australia\\
$^{8}$Australian Astronomical Observatory, PO Box 915, North Ryde, NSW 1670, Australia\\
$^{9}$Jodrell Bank Centre for Astrophysics, School of Physics \& Astronomy, University of Manchester, Oxford Road, Manchester M13 9PL, UK\\
}

\date{Accepted 2016 January 5.  Received 2015 December 31; in original form 2015 November 22.}
\pubyear{2016}

\begin{document}
\label{firstpage}
\pagerange{\pageref{firstpage}--\pageref{lastpage}} 
\maketitle

\begin{abstract}
We investigated the prospects for systematic searches of white dwarfs at low Galactic latitudes, 
using the VLT Survey Telescope (VST) $\Halpha$ Photometric Survey of the Galactic plane and Bulge (VPHAS+). 
We targeted 17 white dwarf candidates along sightlines of known open clusters, aiming to identify
potential cluster members.  We confirmed all the 17 white dwarf candidates from blue/optical spectroscopy, 
and we suggest five of them to be likely cluster members.
We estimated progenitor ages and masses for the candidate cluster members, and 
compared our findings to those for other cluster white dwarfs.
A white dwarf in NGC\,3532 is the most massive known cluster member (1.13\,M$_{\sun}$), 
likely with an oxygen-neon core, for which we estimate an $8.8_{-4.3}^{+1.2}$\,M$_{\sun}$ progenitor, close to the mass-divide between 
white dwarf and neutron star progenitors. A cluster member in Ruprecht\,131 is a magnetic white dwarf, whose progenitor mass
exceeded 2--3\,M$_{\sun}$. We stress that wider searches, and improved cluster
distances and ages derived from data of the ESA {\em Gaia} mission,  
will advance the understanding of the mass-loss processes for low- to intermediate-mass stars.

\end{abstract}

\begin{keywords}
white dwarfs - stars: mass-loss, AGB and post-AGB, neutron - open clusters
\end{keywords}

\section{Introduction}

Main sequence stars of masses below  $\approx 8$--$10\,{\rm M}_{\sun}$
end their lives as white dwarfs \citep[][]{Herwig05,Smartt09}, producing the most
common stellar remnants. Up to 90\,per cent of the mass of white dwarf progenitors is lost 
 on the asymptotic giant branch (AGB), and then dispersed in to the interstellar medium \citep[ISM;][]{Iben83}, 
 enriched with the yields of the nucleosynthesis $s$-processes \citep[][and references therein]{Busso99,Nomoto13}

Quantifying the mass-loss is crucial for a number of reasons. It allows to: 
(i) estimate the amount of stellar yields \citep[e.g.][]{Marigo01, Karakas10,Siess10}, and the dust output 
on the red giant branch (RGB) and AGB \citep[e.g.][]{Matsuura09, McDonald11}; 
(ii) infer the mass-to-light ratio of galaxies \citep[e.g.][]{Maraston98,Kotulla09}; 
(iii) date old stellar populations in open \citep[][]{GarciaBerro10} and globular clusters \citep[][]{Richer97, Hansen04}, 
or in the different constituents of the Milky Way, i.e. the disc \citep[][]{Winget87,Oswalt96}, 
the bulge \citep[][]{Calamida14,Gesicki14}, and the halo \citep[][]{Kalirai12}.
Modelling the final stages of evolution for white dwarf progenitors is complex, 
especially in the super-AGB regime  -- 
that is when 8--10\,M$_{\sun}$ stars could burn carbon under conditions of partial electron degeneracy,
leading either to the formation of stable oxygen-neon core white dwarfs or neutron stars 
via electron-capture supernovae \citep[e.g.][]{Nomoto84,GarciaBerro97, Ritossa99, Farmer15}. 
In this range of masses, the separation between white dwarf and neutron star progenitors is  expected to depend on stellar
properties \citep[metallicity above all;][]{Eldridge04} that influence the mass-growth of the core, as well as the mass-loss 
during the thermally pulsing AGB (TP-AGB) phase \citep[e.g.][]{Siess07, Siess10, Doherty15}. 

Cluster white dwarfs can help to study the correlation between their masses and those of the progenitors, 
known as the initial-to-final mass relation \citep[][]{Weidemann77, Koester80}.
The initial-to-final mass relation can also be studied using white dwarfs in 
wide binaries (i.e. main sequence star plus white dwarf; \citealt{Catalan08b}; 
or two white dwarfs; \citealt{Girven10}; \citealt{Andrews15}), but star clusters
are the most favourable test benches as they can contain samples of white dwarfs,
which formed from a coeval population of stars \citep[][]{PortegiesZwart01}.
Ground-based follow-up spectroscopy is achievable for numerous cluster white dwarfs, 
enabling to assess a wide range of white dwarf progenitor masses, from 1.5--2\,M$_{\sun}$ \citep[e.g.][]{Kalirai08}
to $\gtrsim 7$M\,$_{\sun}$, which are useful to constrain the demarcation between white dwarf and neutron star progenitors \citep[][]{Williams09}.
Although the white dwarf mass distribution is quite well constrained \citep[][and references therein]{Tremblay13}, 
the general trend of the empirical initial-to-final mass relation remains approximate \citep[][]{Weidemann00}, 
especially for low- and high-mass progenitors. Stellar parameters (e.g. metallicity, convection, rotation, magnetic
fields) and environmental effects (e.g. binarity and intra-cluster dynamical interactions) are suggested to add intrinsic scatter to the
shape of the initial-to-final mass relation \citep[e.g.][]{Ferrario05, Catalan08a, Romero15}. 

At present, the study of the cluster initial-to-final mass relation is limited to about 10 clusters, with
$\approx 50$ spectroscopically confirmed white dwarf members \citep[][]{Salaris09}. There are presumably only three open clusters
approaching a fully retrieved white dwarf cooling sequence: the Pleiades \citep[][]{Wegner91, Dobbie06}, the Hyades \citep[][]{Schilbach12} and
Praesepe \citep[][]{Casewell09}, which are all three nearby ($d < 200$\,pc) and above the densest regions
of the Galactic plane ($|b|>10$\,deg). While some new cluster members were discovered
recently \citep[][in NGC\,3532 and M\,37 respectively]{Dobbie12, Cummings15},  
several observational factors have worked against the identification of complete white dwarf populations. First, most clusters
are in crowded, reddened areas of the Galactic plane. Second, the early dispersal of 
clusters causes the number of old clusters to be relatively small \citep[][]{Goodwin06}.
Third, no blue photometric survey, with sufficient magnitude depth ($\lesssim 10$\,mag fainter than 
the cluster turn-off) and angular resolution, covered the Galactic plane until recently.

Here, we test the efficiency of the new VLT Survey Telescope (VST) $\Halpha$ Photometric 
Survey of the Southern Galactic Plane and Bulge \citep[VPHAS+;][]{Drew14} at identifying white dwarfs.
We selected white dwarf candidates in the direction of 11 relatively old open clusters, aiming to confirm new cluster members.
We describe the selection method and observations in Section \ref{chap2}.
The spectral analysis is presented in Section \ref{chap3}, while the estimates of white dwarf parameters and the confirmation of cluster membership are
discussed in Section\,\ref{chap4}. Finally, in Section \ref{chap5}, we derive the progenitor masses for the suggested cluster members,
and compare the new data with initial-to-final mass relations from previous studies. In the concluding remarks, we discuss the future perspectives
for white dwarf searches in the Galactic plane.

\section{The data}
\label{chap2}
\subsection{VPHAS+ photometry}
\label{chap2.1}
\begin{table*}
 \centering
  \caption{Parameters of the 11 open clusters proposed to host the white dwarf candidates.
 We add a tickmark to the last column if at least one new cluster member is identified (Section\,\ref{chap4}).}
\makebox[\textwidth][c]{
  \begin{tabular}{@{}lccrrD{,}{,}{-1}ccccrrrr@{}}
  \hline
   Name & R.A. & Dec & \multicolumn{1}{c}{$\ell$} & \multicolumn{1}{c}{$b$} & \multicolumn{1}{c}{$r_1$,\,$r_2$} & $D^{1,2}$ & $E(B-V)^{1,2}$ & $t_{\rm{oc}}^{1,2}$ & [Fe/H] & $M_{\rm{oc}}^{3}$ & \multicolumn{1}{c}{new} \label{t:clusters}\\
        & (hh:mm:ss) & (dd:mm:ss) & \multicolumn{1}{c}{(deg)}& \multicolumn{1}{c}{(deg)} &\multicolumn{1}{c}{(arcmin)}&(pc) & (mag) & (Myr) &  & $\log(\rm{M}_{\sun})$ &  &\\        
  \hline
  NGC\,2527     & 08:04:58 & $-$28:08:48 & 246.09&$  1.85$ &9,\,20& 601,\,642 & 0.04 & 445,\,800 & -0.10,\,0.20 & 2.5  & \checkmark\\
  BH\,23        & 08:14:24 & $-$36:23:00 & 254.08&$ -0.96$ &9,\,20& 414,\,480 & 0.06 & 250 & & 1.7  & \\
  Platais\,9    & 09:13:47 & $-$43:44:24 & 266.87&$  3.38$ &60,\,126& 174,\,200 & 0.00 & 100 & &  \\
  ASCC\,59      & 10:20:13 & $-$57:39:00 & 283.78&$ -0.51$ &12,\,20& 509,\,550 & 0.05 & 290,\,400 &  &\\
  Loden\,143    & 10:28:54 & $-$58:47:00 & 285.35&$ -0.86$ &10,\,18& 600,\,616 & 0.10,\,0.12 & 281,\,288 & &  &\checkmark\\
  NGC\,3532     & 11:05:39 & $-$58:45:12 & 289.57&$  1.35$ &12,\,25& $492^{4}$ & 0.03,\,0.04 & $300^{5}$ & 0.02 & 2.6  &\checkmark\\
  Platais\,10   & 13:43:28 & $-$59:07:18 & 309.57&$  3.08$ &31,\,60& 246 & 0.00 & 100,\,210 &  &\\
  Johansson\,1  & 15:46:20 & $-$52:22:54 & 327.90&$  1.80$ &11,\,18& 570,\,869 & 0.17 & 200,\,500 & & &\checkmark\\
  ASCC\,83      & 15:50:13 & $-$52:48:00 & 328.10&$  1.11$ &12,\,20& 600,\,619 & 0.12,\,0.15 & 125,\,250 & && \\
  Ruprecht\,131 & 17:49:15 & $-$29:15:00 &   0.14&$ -0.84$ &5,\,10& 600,\,614 & 0.10 & 1480& & 1.1 &\checkmark\\
  Ruprecht\,139 & 18:01:03 & $-$23:32:00 &   6.41&$ -0.24$ &5,\,10& 550,\,593 & 0.10,\,0.15 & 1120& &  \\
\hline
\multicolumn{12}{l}{References: 1) \citet{Dias02}; 2) \citet{Kharchenko13}; 3) \citet{Piskunov08}; 4) \citet{Clem11}}\\
\end{tabular}}
\end{table*}
VPHAS+ started operations in 2011 December 28 and, once completed, will cover the southern Galactic plane between 
$+210^{\circ} \lesssim \ell \lesssim +40^{\circ}$ and $-5^{\circ} < b < 5^{\circ}$, and the Galactic bulge between $|\ell|,\, |b| < 10^{\circ}$.
It combines $ugri$ broad-band filters and a narrow-band $\Halpha$ filter, reaching down to 20\,mag (at 10\,$\sigma$ limit).
The observing strategy of VPHAS+ separately groups (blue) $ugr$ and (red) $r \Halpha i$ frames covering the same field, 
due to different requirements of lunar phase. Therefore, blue and red filters might be observed 
at different epochs. To cover the gaps between CCDs and the cross-shaped shadow cast by the segmented $\Halpha$ filter, 
every $uri$ field is observed at two offset pointings, separated by $-588$\,arcsec and $+660$\,arcsec 
in the RA and declination directions, respectively, while every $g\Halpha$ field is observed at three offset pointings, including an intermediate position.

Here, we use the primary detections of the VPHAS+ data release 2 (DR2), accessible through the ESO Science Archive. 
It delivers PSF magnitudes, expressed in the Vega system, for 24 per cent of the survey area. 
Details on the source detection, photometry, and field-merging are given in the 
data release document\footnote{Available at \url{http://www.eso.org/sci/observing/phase3/data_releases.html}}.
The VPHAS+ DR2 photometry is delivered with a provisional uniform calibration, computed relative to the 
the AAVSO Photometric All-Sky Survey Data Release 8 \citep[APASS;][]{Henden12}, following
the prescriptions given in section 6 of \citet{Drew14}. 
The $u$-band is calibrated separately as explained in section 6 and
fig.\,20 of \citet{Drew14}. 
The zero-points for the $\Halpha$ magnitudes are offset with respect to the $r$-band zero-points,
based on the $(r-\Halpha)$ colours of main sequence stars.
While VPHAS+ DR2 photometry is currently suggested to be consistent with that of the Sloan Digital Sky Survey \citep[SDSS;][]{Abazajian09} 
within 0.05\,mag, there are known systematic errors $\geq 0.1$\,mag in isolated regions of the sky, probably inherited from APASS or due to patchy cloud coverage.
\begin{figure*}
\includegraphics[width=0.85\linewidth]{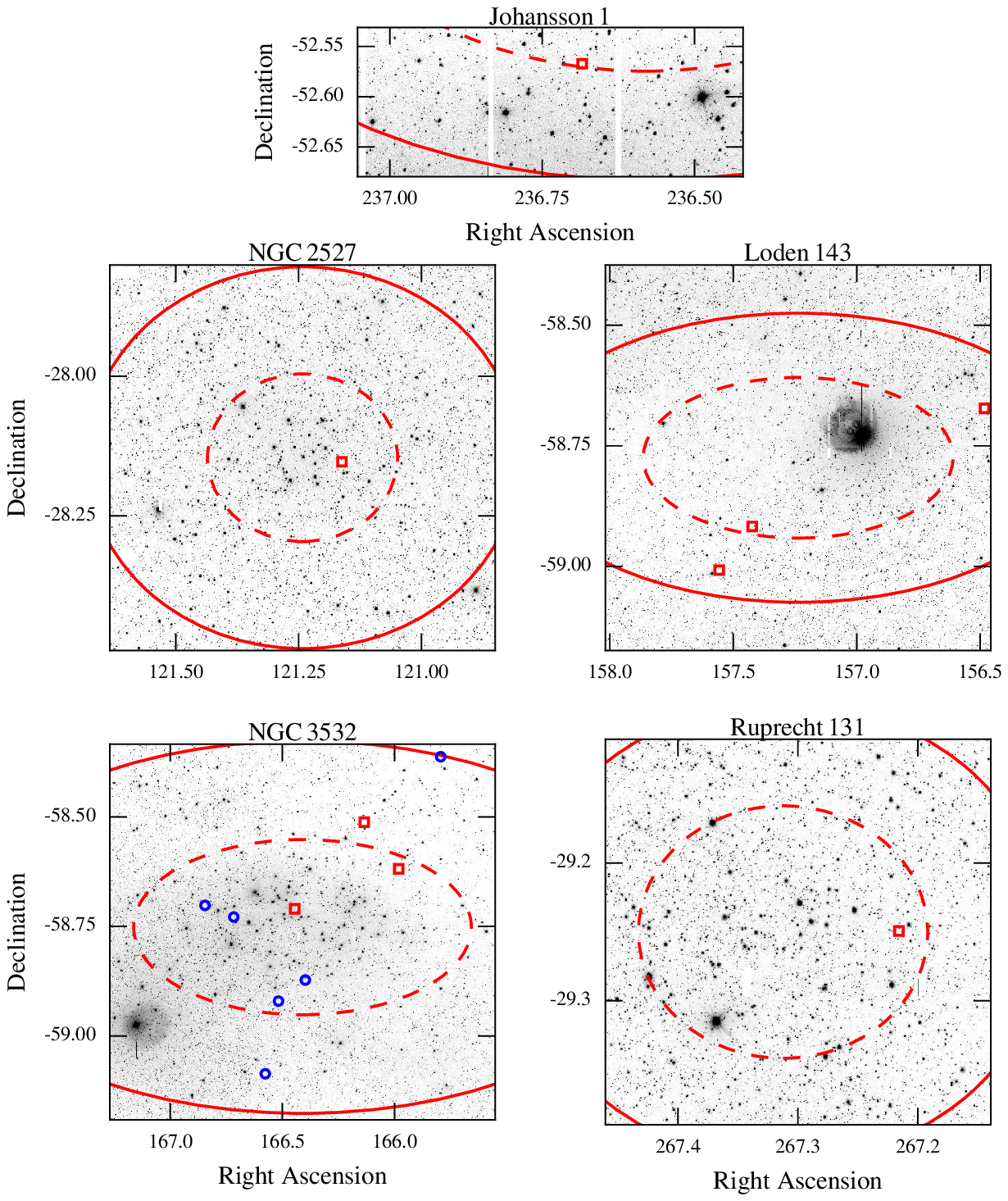}
\caption{VPHAS+ $g$-band mosaics of five open clusters in our sample, for which we identify likely white dwarf members. 
Just a small part of Johanson\,1 is covered by VPHAS+ imaging.
The dashed curves represent the central part of the cluster, $r_{1}$, while the solid curves trace the total cluster area, $r_{2}$ 
\citep[these radii are from][see Table\,\ref{t:clusters}]{Kharchenko13}.
Red squares mark the positions of the white dwarf candidates, blue circles show
the confirmed white dwarfs in NGC\,3532 \citep[][]{Dobbie12}.}
\label{f:charts}
\end{figure*}
\subsection{Clusters}
\label{chap2.2}
The open clusters were drawn from the \citet{Dias02} catalogue,
setting the following criteria:

\begin{itemize}
\item Cluster age $\geq 100$\,Myr, corresponding to the lifetime of a 5\,M$_{\sun}$ white dwarf progenitor
\item Distance modulus $\leq 9.5$\,mag, to have a significant fraction of the white dwarf cooling sequence within the magnitude
limits of VPHAS+. 
\item VPHAS+ $ugr$ photometry covering at least part of the cluster.
\end{itemize}

Of the 45 clusters,  cluster remnants, and stellar associations, which fulfil the first two constraints,  
only 11 currently have VPHAS+ DR2 $ugr$ photometry. 
We list in Table~\ref{t:clusters} their relevant properties, and the bibliographic references.
Distances, reddenings, ages, and metallicities are from \citet{Dias02} and \citet{Kharchenko13}.
The data from \citet{Dias02} are compiled from a number of sources, while \citet{Kharchenko13}
estimated cluster parameters and cluster membership 
using PPMXL \citep[][]{Roeser10} and the Two Micron All Sky Survey \citep[2MASS;][]{Skrutskie06}. 
\citet{Kharchenko13} estimated typical errors
of 11, 7, and 39 per cent for their measures of distances, reddenings, and ages, respectively, 
via comparison of the cluster parameters with data published in the literature. 
In Table~\ref{t:clusters}, we list $r_1$ and $r_2$,
which are the angular radius of the central part and the total radius of the clusters, respectively
\citep[$r_1$ is defined as the angular separation from the cluster centre where the stellar surface density declines abruptly, 
while $r_2$ is the angular separation where the cluster stellar density merges with that of the field;][]{Kharchenko05}.
We note here that Platais\,9 and 10, which have $r_1$ and $r_2$ in the range of 1\,deg,
are presumably stellar associations rather than open clusters, as suggested by \citet{Dias02} and \citet{Kharchenko13}.
The total masses of the clusters, $M_{\rm oc}$, were determined  by \citet{Piskunov08} 
from the inferred tidal radii of the clusters.

The distances of the 11 clusters, given in \citet{Dias02} and \citet{Kharchenko13} 
mostly differ by less than 10\,per cent, with the exception of
Johansson\,1 (570\,pc; \citealt{Dias02}, and 890\,pc; \citealt{Kharchenko13}), probably due to the difficulty 
of determining cluster membership. The cluster ages agree all within $\approx 40$\,per cent. 

In the following paragraphs, we briefly review the available information for five of the 11 open clusters, in which we confirm
new white dwarf members (Section\,\ref{chap4}). In Fig.\,\ref{f:charts}, we show the 
mosaics of VPHAS+ $g$-band frames, covering the central parts of the five clusters, and we mark the positions
of the new white dwarfs we identify as well as those of known white dwarfs.

\subsubsection{NGC\,2527}     
The cluster is well populated with early A-type stars \citep{Lindoff73,Houk75}, which are the brightest stars in Fig.\,\ref{f:charts}.
The range of ages for this cluster (Table\,\ref{t:clusters}) corresponds to turn-off mass of $\approx 2.2$--3.5\,M$_{\sun}$,
i.e. approximately to spectral types A3--B9. 
With a total stellar mass of $\approx 500$\,M$_{\sun}$, we estimated from the \citet{Scalo86} initial mass function that NGC\,2527 
could host $13\pm 6$ white dwarfs. From the VPHAS+ DR2 photometry, we selected for follow-up one white dwarf candidate towards the cluster centre.
\begin{figure*}
\includegraphics[width=\linewidth]{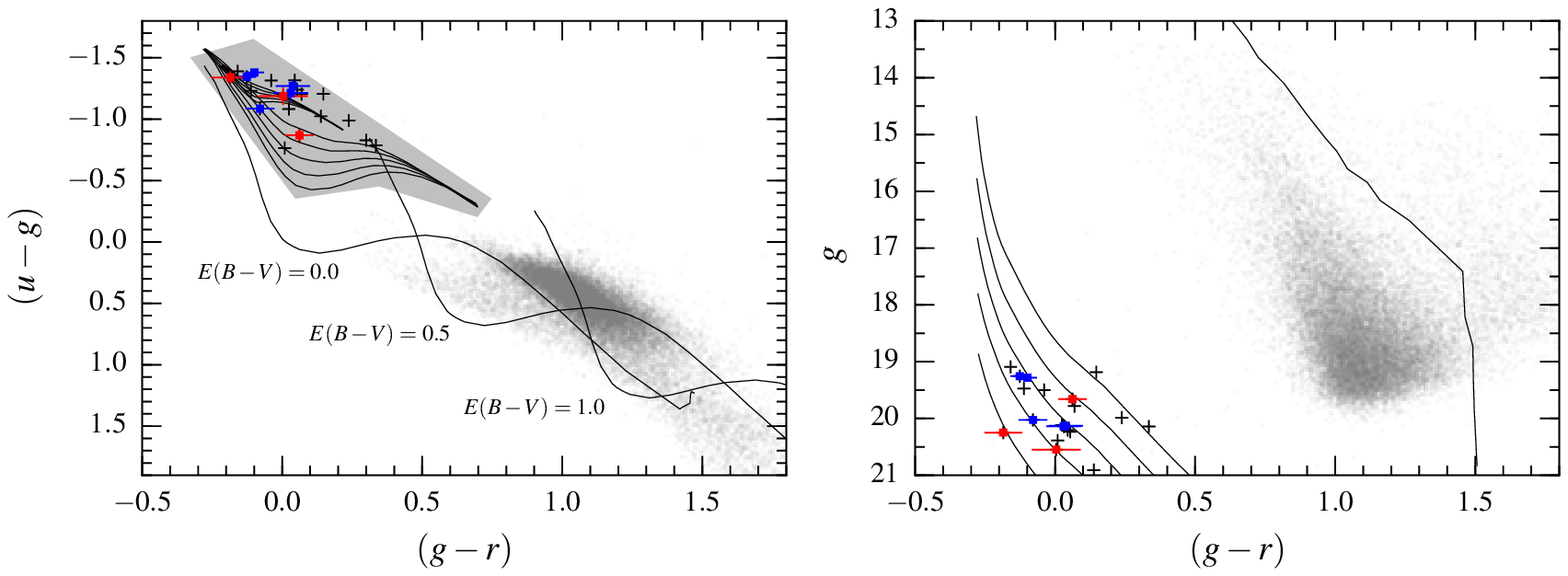}
\caption{Colour-colour (left) and colour-magnitude (right) diagrams 
displaying the point sources identified in the VPHAS+ field 1739, within one cluster radius (18 arcmin) from the centre of NGC\,3532.
On the left, we represent the colour cuts we applied to select white dwarf candidates (black crosses) as a grey shaded area. 
The stars we followed up are overplotted as red squares with errorbars, and the known cluster white dwarfs \citep[][]{Dobbie12}
with VPHAS+ DR2 photometry are represented by blue squares with errorbars. The main sequence \citep[][]{Drew14} is shown as a black curve, which
is displaced by three different magnitudes of interstellar reddening, i.e. $E(B-V) = 0.0,\,0.5,\,1.0$. The white
dwarf tracks for DA and DB white dwarfs (Appendix\,\ref{app:one}) are plotted as black curves,
with $E(B-V) = 0.03$. On the right, in the colour-magnitude diagram,
the main sequence and the DA white dwarf cooling sequence for an assumed cluster distance of 492\,pc, and $E(B-V) = 0.03$ (Table\,\ref{t:clusters}).}
\label{f:ccd_ngc3532}
\end{figure*}
\subsubsection{Loden\,143}	
Like other groups of stars described by \citet{Loden79}, the presence of an open cluster appears in question.
The VPHAS+ frames (Fig.\,\ref{f:charts}) do not show a clearly visible clustering of bright stars.
The very bright, saturated star, which creates an extended reflection in the mosaic of frames,
defines the putative giant branch of the cluster in the colour-magnitude diagrams by \citet{Kharchenko13}.
The authors comment on the sparse appearance of the cluster and suggest its parameters to be poorly constrained.
We observed three white dwarf candidates in the cluster area.
\subsubsection{NGC\,3532}
This is the only cluster in our list with known white dwarf members 
\citep[seven;][]{Reimers89,Koester93,Dobbie09,Dobbie12}. 
In the most recent photometric study of the cluster, \citet{Clem11}
confirmed a distance of $492\pm 12$\,pc and an age of $300\pm100$\,Myr. Using the cluster age, the total mass of the cluster \citep[][]{Piskunov08}, and the initial mass function by \citet{Scalo86},
we expect $\approx 7\pm 4$ white dwarf members. Given the margin for a few more
white dwarfs to be found in this cluster, we followed up three candidates.
Six of the known white dwarfs are shown in Fig.\,\ref{f:charts} (the seventh is outside the figure)
along with the three white dwarf candidates we have observed.  
\subsubsection{Johansson\,1}
The spread in distances and ages reported by \citet{Dias02} and \citet{Kharchenko13}  
is large, and it is probably due to sparse appearance of the cluster that can be also noticed in images from 2MASS and the
VISTA Variables in the Via Lactea (VVV) survey  \citep[][]{Saito12}. Several bright stars
in the cluster area guided the cluster identification by  \citet{Kharchenko13},
but the cluster main sequence was identified mistakenly by \citet{Johansson81} from the study of a few stars in the 
area of  another cluster, Loden\,2326. Unfortunately, VPHAS+ observations do not cover the whole cluster area yet. 
The white dwarf candidates we identified is within the suggested central part of the cluster.
\subsubsection{Ruprecht\,131}      
This cluster is the oldest in our sample and it is found in the Bulge section of the VPHAS+ footprint,
not far from the crowded, young star forming region of the Lagoon nebula.
\citet{Dias02} suggested the identification of Ruprecht\,131 as being dubious. \citet{Kharchenko13} determined its parameters
using the bright stars that are visible in 2MASS images, and also recognisable in the VPHAS+ mosaic of the cluster area.
This cluster is suggested to be old enough for stars down to $\approx 2$\,M$_{\sun}$ to have become white dwarfs (total age < 1.5 Gyr).
Considering the total stellar mass to be less than a hundred solar masses \citep[][]{Piskunov08}, 
it is likely that this cluster has few white dwarfs. We followed up one white dwarf candidate.
\begin{table*}
\scriptsize
  \caption{Details for the 17 white dwarf candidates confirmed by this study, 
  including their photometry with 1\,$\sigma$ errors, 
  VPHAS+ field ID numbers, and observing dates for the red and blue frames.
  The naming convention for VPHAS+ sources is VPHAS\,Jhhmmss.ss+ddmmss.s, which includes the Epoch 2000 coordinates in sexagesimal format.
  In the text, we use an abbreviated version, VPHAS\,Jhhmm+ddmm.}
\makebox[\textwidth][c]{\begin{tabular}{@{}lccccccccc@{}}

  \hline
   Name  & $u$ & $g$ &  $r_{\rm{blue}}$ & $r_{\rm{red}}$ &$\Halpha$ &$i$ &  field & \multicolumn{2}{c}{date-obs}  \label{t:photometry}\\
         & (mag) & (mag) &  (mag) & (mag) & (mag) & (mag) & & (blue) & (red) \\        
  \hline
VPHAS\,J080438.8$-$280914.0  &$19.56 \pm 0.05$ &$20.45 \pm 0.04$&$20.47 \pm 0.08$&$20.45 \pm 0.07$&$20.47 \pm 0.20$&$20.61 \pm 0.15$& 0764 & 2013$-$04$-$13 &2012$-$11$-$13\\
VPHAS\,J081528.4$-$362535.9 &$18.13 \pm 0.02$ &$19.56 \pm 0.02$&$19.66 \pm 0.04$&$19.73 \pm 0.05$&$19.71 \pm 0.08$&$19.70 \pm 0.06$& 0992 & 2013$-$05$-$13 &2012$-$03$-$31\\
VPHAS\,J090004.7$-$455613.4  &$18.53 \pm 0.03$ &$19.92 \pm 0.02$&$20.01 \pm 0.05$&$20.13 \pm 0.08$&&$20.03 \pm 0.13$& 1266 & 2013$-$05$-$02 & 2012$-$04$-$30\\
VPHAS\,J101831.3$-$575211.0  &$19.92 \pm 0.06$ &$20.80 \pm 0.04$&$20.96 \pm 0.11$&$21.07 \pm 0.13$&&$20.64 \pm 0.15$& 1678 & 2012$-$01$-$22 & 2012$-$04$-$29\\
VPHAS\,J102139.0$-$572939.8  &$18.83 \pm 0.03$ &$19.73 \pm 0.02$&$19.63 \pm 0.04$&$19.61 \pm 0.05$&$19.74 \pm 0.09$&$19.58 \pm 0.06$& 1679 & 2012$-$01$-$22 & 2012$-$04$-$29\\
VPHAS\,J102554.7$-$584106.0  &$18.94 \pm 0.04$ &$19.77 \pm 0.03$&$19.59 \pm 0.03$&$19.66 \pm 0.04$&$19.77 \pm 0.10$&$19.54 \pm 0.06$& 1734 & 2012$-$02$-$14 & 2012$-$04$-$29\\
VPHAS\,J102939.4$-$585527.4  &$18.62 \pm 0.02$ &$20.13 \pm 0.04$&$20.25 \pm 0.06$&$20.33 \pm 0.08$&$20.30 \pm 0.15$&$20.29 \pm 0.12$& 1735 & 2012$-$02$-$14 & 2012$-$04$-$29\\
VPHAS\,J103012.0$-$590048.6  &$18.70 \pm 0.03$ &$19.82 \pm 0.03$&$19.86 \pm 0.03$&$19.78 \pm 0.05$&$19.88 \pm 0.10$&$19.92 \pm 0.06$& 1735 & 2012$-$02$-$14 & 2012$-$04$-$29\\
VPHAS\,J110358.0$-$583709.2 &$19.36 \pm 0.05$ &$20.55 \pm 0.05$&$20.54 \pm 0.07$&$20.66 \pm 0.12$&&$20.58 \pm 0.14$& 1739 & 2012$-$02$-$14 & 2012$-$05$-$30\\
VPHAS\,J110434.5$-$583047.4  &$18.91 \pm 0.04$ &$20.25 \pm 0.03$&$20.43 \pm 0.06$&$20.56 \pm 0.10$&$20.17 \pm 0.13$&$20.38 \pm 0.11$& 1739 & 2012$-$02$-$14 & 2012$-$05$-$30\\
VPHAS\,J110547.2$-$584241.8  &$18.79 \pm 0.04$ &$19.66 \pm 0.03$&$19.60 \pm 0.04$&$19.56 \pm 0.05$&$19.93 \pm 0.12$&$19.44 \pm 0.06$& 1739 & 2012$-$02$-$14 & 2012$-$05$-$30\\
VPHAS\,J133741.0$-$612110.2  &$19.16 \pm 0.04$ &$20.36 \pm 0.04$&$20.30 \pm 0.04$&$20.27 \pm 0.05$&$20.30 \pm 0.15$&$20.21 \pm 0.10$& 1900 & 2012$-$02$-$26 & 2012$-$03$-$24\\
VPHAS\,J134436.3$-$613419.2  &$19.37 \pm 0.04$ &$20.71 \pm 0.04$&$20.53 \pm 0.06$&$20.51 \pm 0.06$&$20.78 \pm 0.19$&$20.50 \pm 0.13$& 1900 & 2012$-$02$-$26 & 2012$-$03$-$24\\
VPHAS\,J154644.6$-$523359.0  &$20.18 \pm 0.06$ &$21.20 \pm 0.06$&$21.19 \pm 0.11$&$21.04 \pm 0.10$&$20.84 \pm 0.18$&& 1501 & 2012$-$08$-$14 & 2012$-$07$-$13\\
VPHAS\,J154922.9$-$525158.3  &$19.88 \pm 0.06$ &$20.92 \pm 0.05$&$20.96 \pm 0.11$&$20.94 \pm 0.08$&$20.86 \pm 0.16$&$20.77 \pm 0.13$& 1501 & 2012$-$08$-$14 & 2012$-$07$-$13\\
VPHAS\,J174851.9$-$291456.8  &$20.32 \pm 0.09$ &$21.21 \pm 0.06$&$20.86 \pm 0.08$&$20.84 \pm 0.09$&$20.58 \pm 0.16$&$20.60 \pm 0.18$& 0800 & 2012$-$08$-$10 & 2012$-$06$-$25\\
VPHAS\,J180042.0$-$233238.5  &$17.16 \pm 0.01$ &$18.32 \pm 0.01$&$18.20 \pm 0.02$&$18.30 \pm 0.02$&$18.28 \pm 0.03$&$18.11 \pm 0.03$& 0676 & 2012$-$08$-$14 & 2012$-$06$-$10\\
\hline
\end{tabular}}
\end{table*}
\begin{table*}
 \centering
  \caption{Johnson-Kron-Cousins for the three white dwarfs in NGC\,3532 \citep[][]{Clem11}.}
   \begin{tabular}{@{}lcccc@{}}
  \hline 
name & $B$ & $V$ & $R_{c}$ & $I_{c}$ \label{t:clem11}\\
& (mag)& (mag)& (mag)& (mag) \\
\hline
VPHAS\,J1103$-$5837 & $20.435 \pm 0.024$ & $20.560 \pm 0.027$ & $20.776 \pm 0.081$ & $20.756 \pm 0.660$ \\
VPHAS\,J1104$-$5830 & $20.141 \pm 0.032$ & $20.321 \pm 0.021$ & $20.512 \pm 0.036$ & $20.561 \pm 0.178$ \\
VPHAS\,J1105$-$5842 & $19.713 \pm 0.019$ & $19.581 \pm 0.011$ & $19.639 \pm 0.018$ & $19.615 \pm 0.053$ \\
\hline
\end{tabular}
\end{table*}
\subsection{Photometric selection}
\label{chap2.3}
\begin{table*}
\scriptsize
  \caption{Physical parameters of the 17 white dwarfs confirmed via spectroscopic follow-up. 
  The S/N is measured at $\Hbeta$. $\delta\tau_{\rm WD}$ is the fractional difference in
  cooling ages, between Montreal and {\em BaSTI} models (see Section\,\ref{chap4.1}  for details).}
\makebox[\textwidth][c]{
  \begin{tabular}{@{}llrlD{,}{\pm}{4}cD{.}{.}{6}cD{,}{\pm}{-1}cD{.}{.}{9}D{.}{.}{5}@{}}
  \hline
   Name & cluster &S/N & Type & \multicolumn{1}{c}{$T_{\rm{eff}}$} &  $\log{g}$ & \multicolumn{1}{c}{$M_{g}$} & $E(B-V)$ &  \multicolumn{1}{c}{d} & $M_{\rm{WD}}$ & \multicolumn{1}{c}{$\tau_{\rm{WD}}$} & \multicolumn{1}{c}{$\delta \tau_{\rm{WD}}$}\label{t:physical}\\        
 & & & & \multicolumn{1}{c}{(K)}  &  (cgs) & \multicolumn{1}{c}{(mag)} & (mag) &  \multicolumn{1}{c}{(pc)} & ($M_{\sun}$) & \multicolumn{1}{c}{(Gyr)} & \\
  \hline
VPHAS\,J0804$-$2809 & NGC\,2527     & 20 & DA & 18\,160\,,\,260 &$8.24 \pm 0.05$&11.27_{-0.07}^{+0.09}&$0.05 \pm 0.07$& 630\,,\,77 &$0.77_{-0.03}^{+0.03}$&0.185_{-0.016}^{+0.016} & 0.12\\[1ex]
VPHAS\,J0815$-$3625 & BH\,23	    & 28 & DA & 31\,870\,,\,210 &$8.04 \pm 0.04$& 9.79_{-0.06}^{+0.06}&$0.12 \pm 0.03$& 734\,,\,40 &$0.68_{-0.02}^{+0.02}$&0.011_{-0.002}^{+0.002} & 0.10\\[1ex]
VPHAS\,J0900$-$4556 & Platais\,9    & 23 & DB & 19\,040\,,\,450 &$8.13 \pm 0.10$&10.94_{-0.14}^{+0.14}&$0.01 \pm 0.05$& 615\,,\,59 &$0.70_{-0.06}^{+0.06}$&0.130_{-0.029}^{+0.029} & 0.11\\[1ex]
VPHAS\,J1018$-$5752 & ASCC\,59      & 10 & DA & 17\,430\,,\,390 &$7.79 \pm 0.08$&10.66_{-0.12}^{+0.13}&$0.00 \pm 0.04$&1066\,,\,86 &$0.52_{-0.04}^{+0.04}$&0.095_{-0.010}^{+0.010} & 0.00\\[1ex]
VPHAS\,J1021$-$5732 & ASCC\,59      & 27 & DA & 30\,110\,,\,160 &$8.56 \pm 0.03$&10.79_{-0.05}^{+0.05}&$0.28 \pm 0.04$& 382\,,\,27 &$0.98_{-0.02}^{+0.02}$&0.059_{-0.006}^{+0.006} & 0.10\\[1ex]
VPHAS\,J1025$-$5841 & Loden\,143    & 21 & DA & 13\,280\,,\,240 &$8.44 \pm 0.04$&12.23_{-0.07}^{+0.07}&$0.12 \pm 0.04$& 263\,,\,19 &$0.89_{-0.03}^{+0.03}$&0.580_{-0.028}^{+0.028} & 0.04\\[1ex]
VPHAS\,J1029$-$5855 & Loden\,143    & 14 & DAH & ^{*}35\,000\,,\,5000&$8.00 \pm 0.25$& 9.52_{-0.46}^{+0.46}&$$&1326\,,\,214&$0.66_{-0.11}^{+0.15}$&0.006_{-0.002}^{+0.009} & -0.12\\[1ex]
VPHAS\,J1030$-$5900 & Loden\,143    & 22 & DA & 19\,780\,,\,250 &$8.04 \pm 0.04$&10.81_{-0.06}^{+0.06}&$0.07 \pm 0.04$& 565\,,\,41 &$0.65_{-0.02}^{+0.02}$&0.084_{-0.011}^{+0.011} & 0.00\\[1ex]
VPHAS\,J1103$-$5837 & NGC\,3532     & 19 & DA & 23\,910\,,\,360 &$8.87 \pm 0.05$&11.89_{-0.11}^{+0.11}&$0.13 \pm 0.07$& 433\,,\,54 &$1.13_{-0.03}^{+0.03}$&^{**}0.270_{-0.025}^{+0.034} & \\[1ex]
VPHAS\,J1104$-$5830 & NGC\,3532     & 8  & DC & ^{*}29\,500\,,\,300&$8.00 \pm 0.25$& 9.90_{-0.44}^{+0.44}&$$&1174\,,\,174&$0.64_{-0.11}^{+0.15}$&0.011_{-0.003}^{+0.021} & -0.19\\[1ex]
VPHAS\,J1105$-$5842 & NGC\,3532     & 15 & DA & 14\,230\,,\,1160&$8.16 \pm 0.10$&11.64_{-0.16}^{+0.17}&$0.06 \pm 0.04$& 363\,,\,36 &$0.71_{-0.06}^{+0.06}$&0.323_{-0.066}^{+0.086} & 0.11\\[1ex]
VPHAS\,J1337$-$6121 & Platais\,10   & 14 & DA & 24\,100\,,\,530 &$8.01 \pm 0.08$&10.37_{-0.13}^{+0.13}&$0.20 \pm 0.05$& 709\,,\,67 &$0.64_{-0.04}^{+0.05}$&0.032_{-0.005}^{+0.013} & 0.00\\[1ex]
VPHAS\,J1344$-$6134 & Platais\,10   & 10 & DA & 49\,670\,,\,2520&$7.92 \pm 0.19$& 8.87_{-0.37}^{+0.37}&$0.39 \pm 0.06$&1202\,,\,169&$0.66_{-0.08}^{+0.09}$&0.002		    & -0.81\\[1ex]
VPHAS\,J1546$-$5233 & Johansson\,1  & 12 & DA & 20\,440\,,\,550 &$7.98 \pm 0.09$&10.65_{-0.15}^{+0.15}&$0.12 \pm 0.11$&1052\,,\,204&$0.62_{-0.04}^{+0.05}$&0.063_{-0.006}^{+0.019} & -0.01\\[1ex]
VPHAS\,J1549$-$5251 & ASCC\,83      & 14 & DA & 20\,320\,,\,390 &$8.05 \pm 0.07$&10.77_{-0.11}^{+0.11}&$0.07 \pm 0.10$& 951\,,\,165&$0.66_{-0.04}^{+0.04}$&0.078_{-0.014}^{+0.017} & 0.01\\[1ex]
VPHAS\,J1748$-$2914 & Ruprecht\,131 & 12 & DAH & ^{*}8750\,,\,1200&$8.00 \pm 0.25$&12.84_{-0.52}^{+0.61}&& 471\,,\,102&$0.60_{-0.12}^{+0.16}$&0.854_{-0.243}^{+0.672} & 0.05\\[1ex]
VPHAS\,J1800$-$2332 & Ruprecht\,139 & 21 & DA & 24\,030\,,\,360 &$8.60 \pm 0.05$&11.36_{-0.09}^{+0.09}&$0.24 \pm 0.02$& 165\,,\,7  &$0.99_{-0.03}^{+0.03}$&0.153_{-0.019}^{+0.019} & 0.21\\[1ex]
\hline
\multicolumn{9}{l}{*: Effective temperatures of DAH and DC white dwarfs are estimated from photometric fitting; no E(B-V) estimates.}\\
\multicolumn{9}{l}{**: Cooling age determined from \citet{Althaus07} oxygen-neon core models.}
\end{tabular}}
\end{table*}

White dwarfs occupy a limited part of the $(u-g,\,g-r)$ colour plane that is also populated by hot subdwarfs, O- and B-type stars, and quasars
\citep[e.g.][]{Girven11,Greiss12,Verbeek12}, whose contamination can be efficiently suppressed by applying reduced proper-motion selection criteria
\citep[][]{GentileFusillo15}. Furthermore, in the Galactic plane, the
contamination by quasars is expected to be insignificant, due to the
blocking effect of the interstellar reddening. Hot subdwarfs and high-mass main sequence stars have redder colours than white dwarfs, 
because they are more distant \citep[e.g.][]{Mohr-Smith15}.

We identified about 70 white dwarf candidates, towards the 11 selected clusters, via cuts in the $(u-g,\,g-r)$ colour-colour diagram, 
based on the synthetic colours of hydrogen- (DA) and helium-dominated (DB) white dwarfs (see example in Fig.\,\ref{f:ccd_ngc3532}).
To guide our selection, we corrected the white dwarf tracks according to the interstellar reddening of the open clusters (Table\,\ref{t:clusters}). 
The absolute magnitudes of white dwarfs were computed in the VPHAS+ Vega system, 
convolving the transmission curves of the filters with a grid of \citet{Koester10} synthetic spectra.
The fluxes of model spectra were calibrated to an absolute scale following \citet{Holberg06}, and using the mass-radius relation adopted by the 
Montreal group\footnote{Available at: \url{http://www.astro.umontreal.ca/~bergeron/CoolingModels}}. 
The absolute $g$-band magnitudes and intrinsic colours in the VPHAS+ Vega system are given in Appendix\,\ref{app:one}, for a range of 
atmospheric temperatures ($T_{\rm eff} = 6000$--$100\,000$\,K) and surface gravities ($\log{g} = 7$--9\,dex).

To maximise the chance of identifying cluster members and to prioritise the targets for the spectroscopic follow-up,
 we estimated photometric distances and cooling ages of the white dwarf candidates.
Using DA models at fixed $\log{g} = 8$,  we estimated the $T_{\rm{eff}}$ by fitting the VPHAS+ $ugr$ photometry. Next, 
 we inferred the absolute magnitudes of the white dwarf candidates interpolating the tables in Appendix\,\ref{app:one},
 and we estimated their cooling ages from the cooling models of the Montreal group \citep[][]{Fontaine01}. Finally, we estimated their photometric parallaxes.  
We chose to follow up 17 targets (see next section), having
 photometric distances and cooling ages broadly consistent
with those of the selected open clusters.
We summarise the relevant VPHAS+ data for the 17 spectroscopic targets in Table~\ref{t:photometry}.

Three white dwarf candidates in the area of NGC\,3532 
also have Johnson-Kron-Cousins $BVR_{c}I_{c}$ photometry \citep[][]{Clem11}, listed in Table\,\ref{t:clem11}.
The $B$ and $V$ magnitudes are in good agreement with VPHAS+ DR2 photometry. 
The $R_{c}$ and $I_{c}$ magnitudes carry larger errors, but they
appear to hint at small systematic differences with VPHAS+ DR2 at the faintest magnitudes.

\subsection{Optical spectra}
\label{chap2.4}

We acquired optical spectroscopy for 17 white dwarf candidates on 2014 April 28--30 
with the visual and near-UV FOcal Reducer and low dispersion Spectrograph \citep[FORS2;][]{Appenzeller98}, 
mounted on the Very Large Telescope (VLT) UT1 (Antu). 
We used the blue sensitive E2V CCDs, with a pixel size of 15\,$\mu$m, and the Grism 600B+22, 
which give a dispersion of 50\,\AA/mm. The SR collimator (f1233\,mm) was used
with the standard $2\times2$ binned readout mode, giving a plate scale of 0.25\,arcsec. With a 0.7\,arcsec wide slit, 
we obtained a resolving power of $R \approx 1000$ at $\Hbeta$.
The relevant spectral coverage is 3500--6100\,\AA, allowing to cover all the Balmer series from $\Hbeta$ to the Balmer jump.

Weather conditions were overall good, but not photometric, with seeing varying between 0.5--1.6\,arcsec. 
The exposure times ranged over 300--1200\,s, and we achieved a signal-to-noise ratio (S/N) of
 $\geq 10$ at $\Hbeta$ for most stars (see Table~\ref{t:physical}). 
One spectrophotometric standard was observed each night, to allow for relative flux-calibration. Standard calibrations were
taken at the end of the night (bias, flat-fields, HeAr arc lamps).

The 2D images were reduced in a standard fashion to remove the bias, to apply flat-field correction and wavelength calibration, 
to extract the 1D spectrum, and to apply the flux calibration. 
The reduction steps were undertaken with the software developed by T.\,R.\,Marsh, {\sc pamela} \citep{Marsh89}
 and {\sc molly}\footnote{{\sc pamela} is part of the {\sc starlink} distribution at
 \url{http://starlink.eao.hawaii.edu/starlink}.
{\sc molly} is available at \url{http://www.warwick.ac.uk/go/trmarsh/software/}.}.
The extracted spectra are shown in Fig.~\ref{f:all_fits}.
Due to the relatively large sky background and faint magnitudes of the targets, 
some of the extracted spectra show residual sky-lines at 5577\,\AA. 

The flux calibrated spectra follow relatively well the slope of the VPHAS+ DR2 $ugr$ photometry, although 
some slight differences are apparent in the $u$-band. 
Since our observations do not extend below 3500\,\AA, we cannot fully
determine the flux contribution to the $u$-band from the observed spectra.
For one object, VPHAS\,J1021$-$5732, the slope inferred from the photometry 
appears to be $\approx 0.15$\,mag redder than that of the VLT/FORS2 spectrum.
As it remains unclear whether it is a problem in the flux calibration of the spectrum 
or it is related to the DR2 photometry,
the distance determined in Section\,\ref{chap4} could be affected.    

\section{Spectral analysis}
\label{chap3}
\begin{figure}
\includegraphics[width=\linewidth]{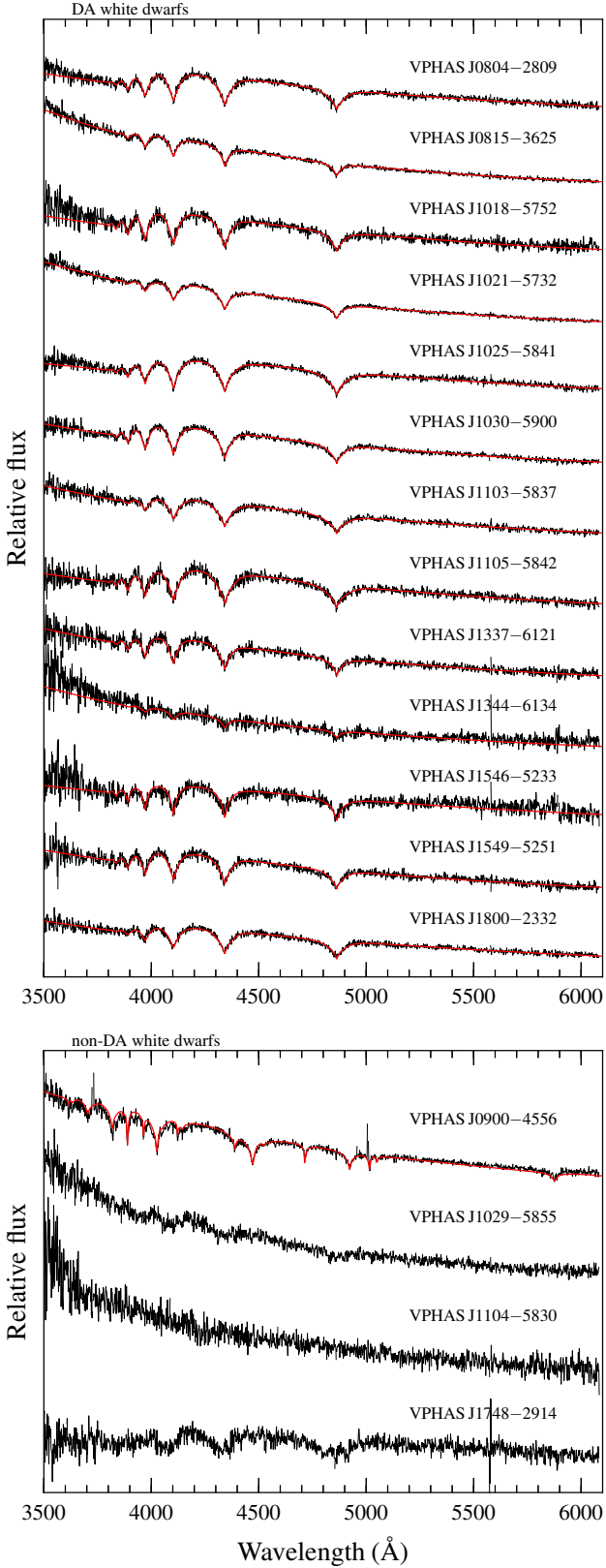}
\caption{Top panel: in black, the observed spectra of DA white dwarfs (top panel) and, in red,
the best-fitting models, normalised to the observed spectra. Bottom panel: as before, from top to bottom, 
the DB, DAH, DC, and DAH white dwarfs. The mismatch between some observed and 
model spectra at $\approx 3600$\,\AA\ is likely due to calibration issues.}
\label{f:all_fits}
\end{figure}
\begin{figure}
\includegraphics[width=\linewidth]{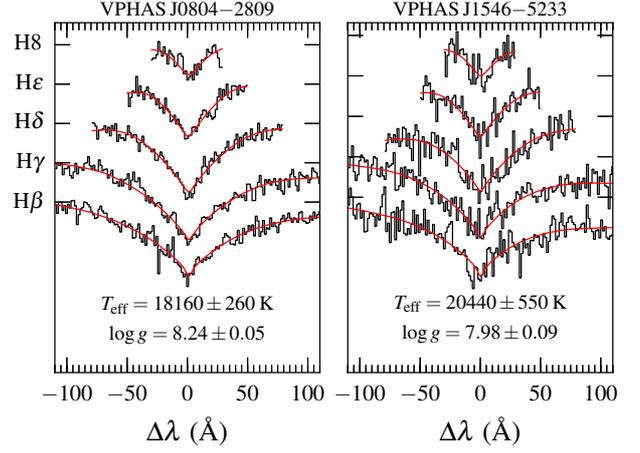}
\caption{Normalised Balmer-line profiles for two spectra in our sample (black),
and best-fit model spectra (red).}
\label{f:example_fit}
\end{figure}
We confirmed all 17 targets to be white dwarfs.
Inspection of the spectra (Fig.~\ref{f:all_fits}) reveals 13 hydrogen-line (DA) white dwarfs, one helium-line (DB) white dwarf (VPHAS\,J0900$-$4556), 
two likely magnetic (DAH) white dwarfs (VPHAS\,J1029$-$5855, VPHAS\,J1748$-$2914) with visible
Zeeman splitting of the hydrogen lines, and a continuum (DC) white dwarf (VPHAS\,J1104$-$5830). 

The atmospheric parameters ($T_{\rm{eff}}$, $\log{g}$) of the DA white dwarfs were determined via comparison with two sets of model spectra.
First, we used the {\sc fitsb2} program \citep[][]{Napiwotzki04} with \citet{Koester10} model spectra. 
{\sc fitsb2} performs a fit to the spectral lines, minimising the $\chi^{2}$
with a downhill simplex algorithm \citep[derived from the AMOEBA routine;][]{Press92}.
The adopted grid of synthetic spectra spans $T_{\rm{eff}} = 6\,000$--100\,000\,K and $\log{g} = 5$--9.
The errors were assessed via a bootstrap method.
Second, we used the set of model spectra by \citet[][]{Tremblay11} and followed the fitting procedure detailed for DA white dwarfs in \citet{bergeron92}.
Both the \citet{Koester10} and \citet{Tremblay11} model spectra implement improved Stark broadening profiles of the hydrogen lines computed by
\citet{Tremblay09}. The \citet[][]{Tremblay11} models also account for NLTE effects,
more appropriate for the study of hot white dwarfs like VPHAS\,J1344$-$6134.
Here we adopted for both sets of model atmospheres the mixing-length prescription ML2/$\alpha = 0.8$.
Due to the inaccurate treatment of convection, parametrised by the mixing-length theory in 1D model atmospheres, 
the $\log{g}$ measured from line-profile fits tends to be overestimated. Therefore,
following \citet{Tremblay13}, we corrected the measured $T_{\rm eff}$ and $\log{g}$ of DA white dwarfs with
$T_{\rm eff} < 15\,000\,K$.

We measured the atmospheric parameters, using the following five transitions: $\Hbeta$, 4861.3\,\AA;
H$\delta$, 4340.5\,\AA; H$\gamma$, 4101.7\,\AA; H$\epsilon$, 3970.4\,\AA; H8, 3889.05\,\AA.
We achieved an accuracy of 200--850\,K and 0.06-0.13, for the estimates of $T_{\rm eff}$ and $\log{g}$,
with reduced $\chi^{2}$ of the order of unity. Given that the atmospheric parameters we measured using the two grids of models 
agreed to better than $2\,\sigma$ in all cases, we adopted the average values. To illustrate the quality of the data,
we display the normalised Balmer lines of two observed spectra, VPHAS\,J0804$-$2809 and VPHAS\,J1546$-$5233, 
and the corresponding best-fit model spectra, in Fig.\,\ref{f:example_fit}. 
In some cases, the fitting procedure led to two possible solutions due to a degeneracy 
between $T_{\rm eff}$ and $\log{g}$, namely the hot and cool solutions. We compared the results with the observed photometry in order to choose
the most likely correct solution. 

For some of the noisiest spectra, with S/N\,$\leq$\,20, the line profiles appear
distorted and could arise from the superposition of two DA white dwarfs. Due to the quality of the data and the wavelength coverage of our spectra, 
we cannot rule out the presence of unseen, close white dwarf companions, suggested to be $\approx 25$\,per cent of the 
field population \citep[][]{Nelemans01}, or more in old open clusters \citep[][]{PortegiesZwart01}.
Existing near-infrared data seem to exclude the presence of low-mass late-type companions (see next section for further discussion).  

To determine the atmospheric parameters of VPHAS\,J0900$-$4556, we used {\sc fitsb2} with \citet{Koester10} DB model spectra, fitting 
the following He\,{\sc i} lines: 4921.9, 4713.1, 4471.5, 4026.2, and 3888.7\,\AA.
The DB grid of spectra spans $T_{\rm{eff}} = 10\,000$--40\,000\,K and $\log{g} = 7$--9. 
For the two DAH and the DC white dwarfs, we estimated photometric $T_{\rm eff}$ from the available VPHAS+ DR2 magnitudes, 
with DA and DB model spectra, respectively. This precluded a determination of $\log{g}$ and interstellar reddening. 
To assess the $T_{\rm eff}$ uncertainty, we considered $\log{g} = 8.00 \pm 0.25$\,dex, 
corresponding to a white dwarf mass of $0.60 \pm 0.15$\,M$_{\sun}$. 
Given the high $T_{\rm eff}$ we estimated for VPHAS\,J1104$-$5830, which is anomalous for typical DC white dwarfs, 
we suspect this star to be also magnetic. In presence of strong magnetic fields, 
the energy levels of the dominant atmospheric elements are characterised by large shifts, 
which would make the low S/N spectrum of VPHAS\,J1546$-$5233 look featureless.

We list the atmospheric parameters of all the observed white dwarfs in Table~\ref{t:physical} and
we overplot the corresponding model spectra on the VLT data in Fig.~\ref{f:all_fits}.

\subsection{Interstellar reddening}
\label{chap3.1}
\begin{figure*}
\includegraphics[width=0.95\textwidth]{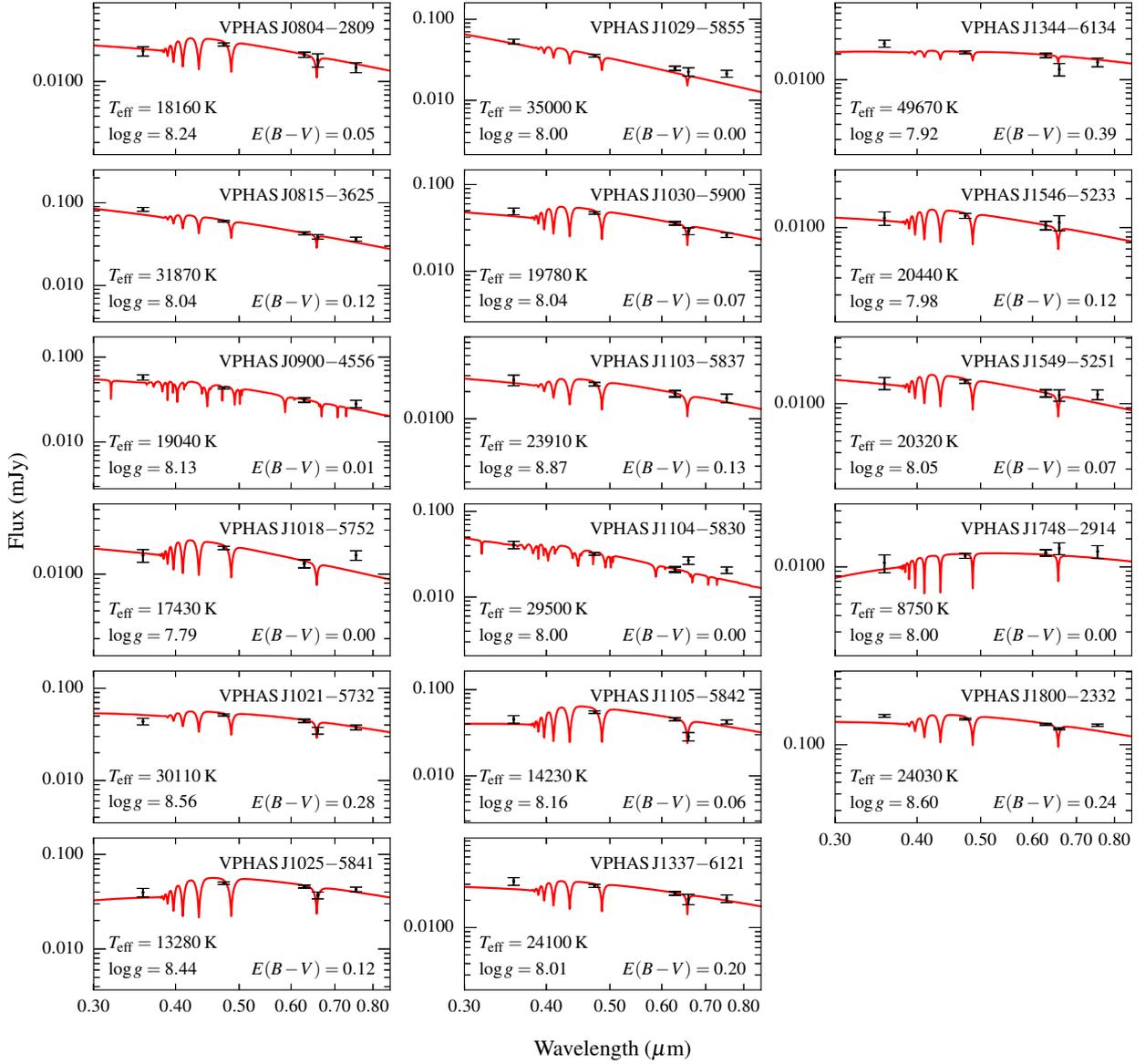}
\caption{Observed fluxes (black errorbars) and best-fit models to the spectral lines (red). 
Model parameters and interstellar reddening are indicated in each panel.
The model spectra of DA white dwarfs are reddened to match the $(g-r)$ colour. 
DB models are plotted for VPHAS\,J0900$-$4556, and the DC white dwarf, VPHAS\,J1104$-$5830, 
whose low S/N spectrum does not reveal any noticeable spectral line.
DA models are also plotted for the two DAH white dwarfs, 
VPHAS\,J1029$-$5855 and VPHAS\,J1748$-$2914, whose $T_{\rm eff}$ are estimated from
photometric fit, keeping $\log{g} = 8$ and $E(B-V) = 0$.}
\label{f:photometry}
\end{figure*}
\begin{figure}
\includegraphics[width=\linewidth]{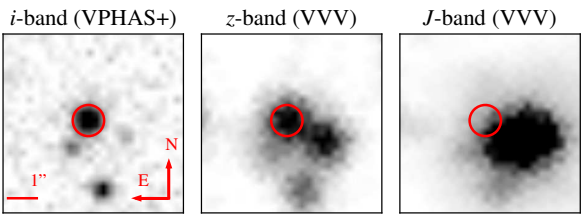}
\caption{Image cut-outs of VPHAS\,J1800$-$2332. 
The infrared flux in the $J$-band is likely associated to another star at 1.25\,arcsec from VPHAS\,J1800$-$2332, 
which is just visible in $i$-band frame, but it is already bright in the $z$-band.}
\label{f:vphasj1800_images}
\end{figure}
\begin{figure*}
\includegraphics[width=0.95\linewidth]{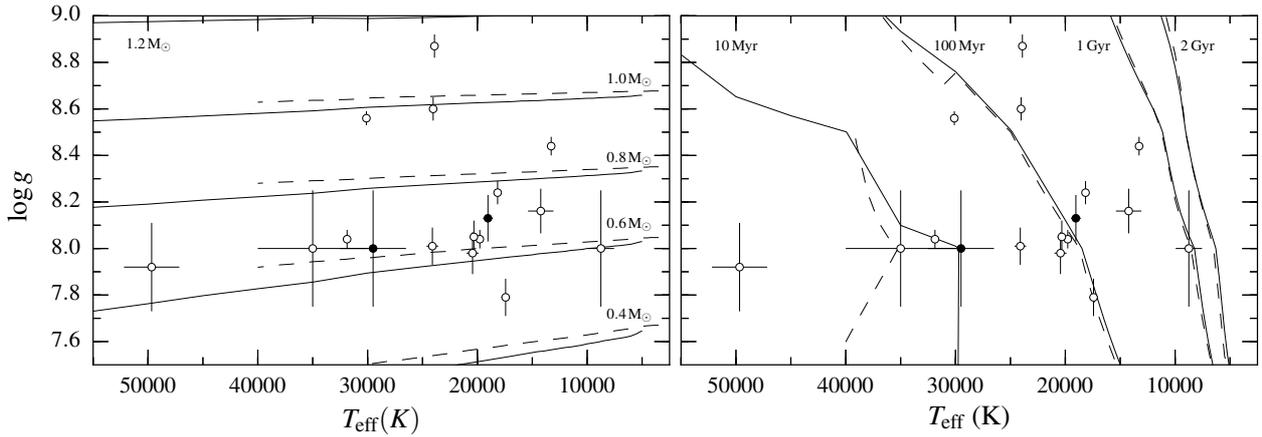}
\caption{Mass (left panel) and ages (right panel) are estimated via interpolation of
$T_{\rm eff}$ and $\log{g}$ on to the white dwarf cooling tracks of \citet{Fontaine01}. 
DA tracks are plotted as solid lines, while dashed curves represent DB tracks. 
The filled dots are the DB and DC white dwarfs, i.e. VPHAS\,J0900$-$4556 and VPHAS\,J1104$-$5830.}
\label{f:parameters}
\end{figure*}
\begin{figure}
\includegraphics[width=\linewidth]{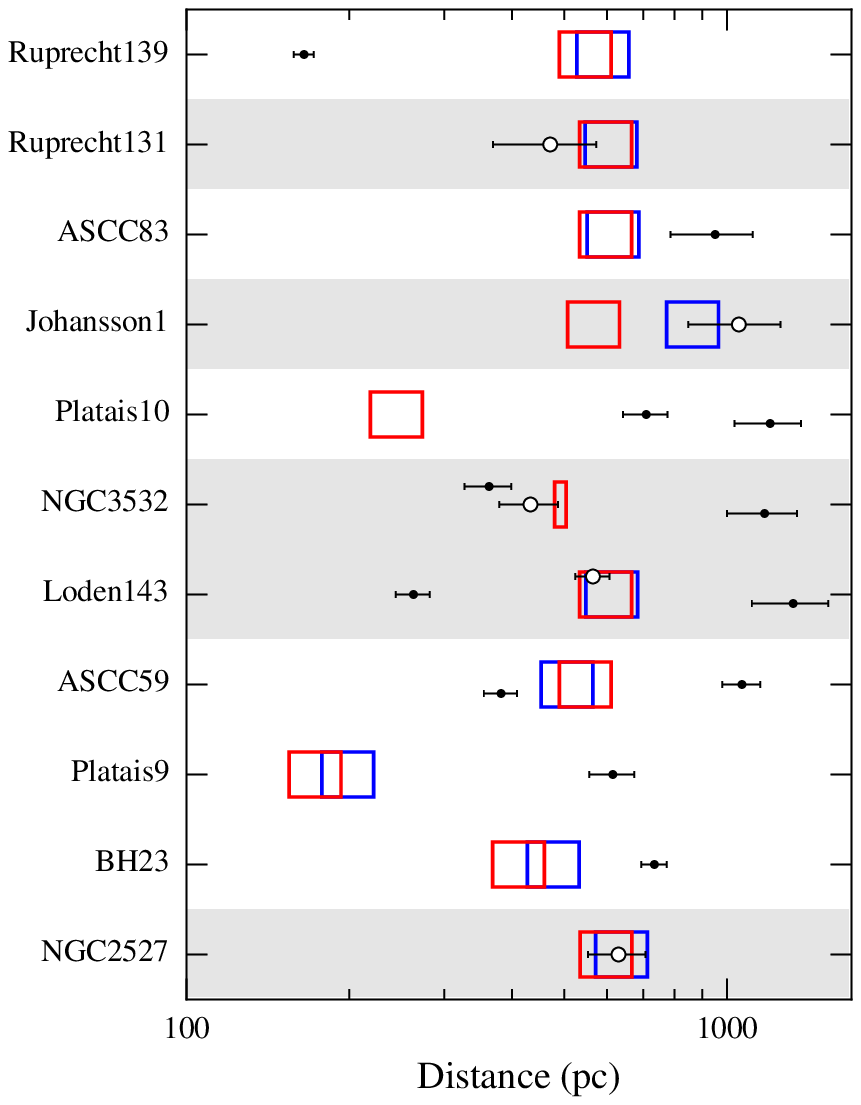}
\caption{Assessment of cluster membership for the 17 white dwarfs. 
Red boxes indicate cluster distances of \citet{Dias02}, 
blue boxes those of \citet{Kharchenko13}, and points with error bars the white dwarf distances. 
The box sizes correspond to an uncertainty of 11\,per cent of the cluster distance \citep[][]{Kharchenko13}. 
The candidate cluster members are represented by white circles with error bars, 
and the corresponding rows are highlighted by a grey shaded area.}
\label{f:distances}
\end{figure}

The comparison between the intrinsic and observed colours of the 17 white dwarfs suggests
modest amounts of interstellar reddening.
We determined the colour excess in the $(g-r)$ colour as:
\begin{equation}
E(B-V) = 0.86\times E(g-r) = 0.86\times[(g-r) - (g-r)_{\circ}],
\end{equation}
where the $r$ magnitudes we used are the $r_{\rm blue}$ in Table\,\ref{t:photometry}.
The conversion factor between $E(B-V)$ and $E(g-r)$, is derived from the standard $R_{V} = 3.1$ reddening
law by \citet{Fitzpatrick99}. The intrinsic colour, $(g-r)_{\circ}$, is interpolated from the tables in the Appendix\,\ref{app:one}, 
at the corresponding $T_{\rm eff}$ and $\log{g}$ of each star.

In Fig.\,\ref{f:photometry}, we display the model spectra (derived either from spectroscopic or photometric fit) 
of the 17 white dwarfs along with the observed photometry, while the measured reddenings are given in Table~\ref{t:physical}. 
The agreement between photometry and model atmospheres is overall good, although 
some disagreement is seen in the $u$-band, which is calibrated following the prescriptions given by \citet{Drew14}. 
The calibration may be problematic in some reddened field, due to the sparse appearance of the colour-colour diagram. 
Thus, the $u$ magnitudes carry larger systematic uncertainties, because 
their calibration depends on that of $g$ and $r$ magnitudes, and it is also more subject to variations of atmospheric transparency.

The $i$-band fluxes of a few objects appear to be slightly above the predictions given by the model atmospheres.
Thus, we checked if the observed white dwarfs display excess  
in the near-infrared, i.e. 2MASS, WISE \citep[][]{Wright10}, and VVV, signalling the presence of low-mass late-type companions.
We note a bright infrared source at 1.25\,arcsec from VPHAS\,J1800$-$2332 (Fig\,\ref{f:vphasj1800_images}). 
The flux from this object is not likely to affect the $i$ magnitudes of the white dwarf, therefore we suspect that in this and 
other cases the calibration might suffer with systematically larger offsets, arising from the APASS-based  calibration.

For VPHAS\,J1103$-$5837 and VPHAS\,J1105$-$5842, in NGC\,3532, we also estimated the interstellar reddening from their $(B-V)$ colours
\citep[][Table\,\ref{t:clem11}]{Clem11}, obtaining $0.07\pm 0.04$ and $0.08\pm 0.02$\,mag, respectively, compatible with those we measure from VPHAS+ data.

\section{Masses, cooling ages, and distances}
\label{chap4}
To establish the cluster membership of white dwarfs, we need to estimate their distances
and verify that the cooling ages are compatible with the cluster ages. 
 
We determined the white dwarf spectroscopic parallaxes, using the appropriate intrinsic 
magnitudes from Appendix\,\ref{app:one}, the observed magnitudes, and the interstellar extinction as $A_{g} = 3.68 \times E(B-V)$. 
The absolute magnitudes and distances are given in Table\,\ref{t:physical}.
Then, we estimated white dwarf masses and cooling ages from the cooling tracks of the Montreal group
\citep[][see Fig.\,\ref{f:parameters} and Table\,\ref{t:physical}]{Fontaine01}. 
For DA white dwarfs, we used the cooling models with thick 
hydrogen atmospheres ($10^{-4}$\,M$_{\sun}$) and carbon-oxygen cores \citep[][]{Bergeron01}.
Above $30\,000$\,K, the carbon-core cooling models by  \citet{Wood95} were used instead. 
For DB white dwarfs, we used the cooling models with a thinner hydrogen layer of $10^{-10}$\,M$_{\sun}$.

 \citet{Salaris09} assessed the effect of systematic differences introduced by
different treatments of neutrino cooling, core composition, and envelope thickness, for their cooling models. 
Referring to their table\,4, we found that an increased neutrino cooling rate
would produce a difference of 7--34\,per cent in the white dwarf cooling ages of our sample, 
depending on $T_{\rm eff}$ and $\log{g}$. Smaller uncertainties are derived for different
conductive opacities, core composition, and hydrogen-layer thickness, of the order of 2--6\,per cent. We took these uncertainties 
into account when determining the progenitor lifetimes and masses, in the following section.
In order to assess the effect of different cooling tracks on the age estimates, we also
computed white dwarf cooling ages using the {\em BaSTI} models \citep[][]{Salaris10},
which use different formulations with respect to those of \citet{Fontaine01} for the 
 equation of state and opacities. In Table\,\ref{t:physical}, 
we list the fractional difference in cooling ages, expressed as 
$\delta \tau_{\rm{WD}}  = [\tau_{\rm{WD}} ({\rm Montreal})-\tau_{\rm{WD}} (BaSTI)] / \tau_{\rm{WD}} ({\rm Montreal})$. 
The effect is comparable to the other uncertainties, and it is mostly in the range of 0.10\,dex.

One white dwarf, VPHAS\,J1103$-$5837, could have
an oxygen-neon core \citep[$M > 1.06$\,M$_{\sun}$;][]{GarciaBerro97}. 
Thus, we used the \citet{Althaus07} cooling models for oxygen-neon cores, 
which suggest a cooling age $\approx 10$\,per cent  shorter than that of a carbon-oxygen core white dwarf with the same mass. 

\subsection{Cluster membership}
\label{chap4.1}
Comparing the white dwarf distances (Table~\ref{t:physical}) with the cluster distances (Table\,\ref{t:clusters}) in Fig.\,\ref{f:distances},
we find that five of the 17 white dwarfs overlap within 1\,$\sigma$ with the putative clusters (Table\,\ref{t:progenitor}).
The five white dwarfs have cooling ages younger than the cluster ages,
which is also a necessary requirement for cluster membership. 

Our sample of photometrically selected white dwarfs is dominated by the field population, given that 
the preliminary identification (Section\,\ref{chap2.3}) does not allow to estimate accurate parallaxes.  
The inclusion of proper motions is a valuable tool, 
which can be considered in future for discriminating with higher accuracy between field and cluster members,
although there may not be available data for the faint white dwarf studied here.

In order to estimate the contamination of field white dwarfs at the distance of each cluster,
we used the white dwarf luminosity function derived from SDSS \citep[see fig.\,4 in][]{Harris06}, which gives a space density of 
0.0046 white dwarfs per pc$^{-3}$. Since the luminosity function expresses the space density of white dwarfs 
in function of their bolometric magnitudes, we converted it to an apparent magnitude scale 
using the distances and the reddenings of the five open clusters, for which we identify white dwarf member candidates.
Then, we integrated the luminosity functions between the range of apparent magnitudes for white dwarfs with cooling ages compatible to those of the clusters,
providing they are within the VPHAS+ magnitude limits ($13 \leq g \leq 22$).
Given the  extension of the five open clusters ($\approx 0.125$\,deg$^{2}$),
we expect $\leq 1$--2 field white dwarfs within the angular radius $r_2$ of each cluster (Fig.\,\ref{f:charts}). 
We note that this number is small with respect to the number of expected white dwarfs
in old clusters of 500--1000\,M$_{\sun}$, therefore we consider the contamination by field white dwarfs to be negligible.

Four stars, including three cluster members, deserve further mention. The first is VPHAS\,J1103$-$5837, in NGC\,3532. 
The interstellar reddening we measure for this white dwarf from VPHAS+ DR2 colours, $E(B-V) = 0.13 \pm 0.07$, 
is slightly larger ($2\,\sigma$) than that of the cluster (Table\,\ref{t:clusters}). Using the \citet{Clem11} photometry in Table\,\ref{t:clem11},
we measure $E(B-V) = 0.07 \pm 0.04$, enabling a better comparison with the cluster reddening. These small differences in interstellar reddening
do not modify much the white dwarf distance, and its location within the central part of the cluster (dashed curve in Fig.\,\ref{f:charts}) 
adds further evidence that this massive white dwarf may belong to NGC\,3532. 
Second, for VPHAS\,J1546$-$5233 in Johansson\,1, we measured distance and reddening that are compatible with those measured by
\cite{Kharchenko13}, therefore we used their cluster age to infer the progenitor parameters for this white dwarf in the following section.
Third, for VPHAS\,J1748$-$2914, which is a magnetic white dwarf in Ruprecht\,131, we estimated the $T_{\rm eff}$ via
a photometric fit, assuming a value of $\log{g} = 8.00 \pm 0.25$ based on the typical mass distribution of field white dwarfs \citep[][]{Tremblay13}.
 Since magnetic white dwarfs are often suggested to be slightly more massive 
than non-magnetic white dwarfs \citep[][and references therein]{Ferrario15},
they are more compact and less luminous at a given $T_{\rm eff}$.
This implies that VPHAS\,J1748$-$2914 could be at a shorter distance, which may not be compatible with the cluster distance.
Thus, the association of this white dwarf to Ruprecht\,131 needs a stronger confirmation, via higher quality spectroscopy allowing
more precise typing. Finally, VPHAS\,J1104$-$5830 that we also speculated to be a magnetic white dwarf in Section\,\ref{chap3},
could be closer than 1174\,pc. However, it is worth noting that, even for $\log{g} = 8.5$, 
corresponding to $M \approx 0.94$\,M$_{\sun}$,
its distance would be $\approx 800$\,pc, which is still further away than NGC\,3532.

\section{Discussion}
\label{chap5}
\subsection{Progenitor ages and masses}
\label{chap5.1}
\begin{table*}
 \centering
  \caption{Physical parameters of the white dwarf progenitors for the five likely cluster members, 
  determined via interpolation of the progenitor ages with the {\em BaSTI} and
  \citet{Ekstrom12} isochrones for rotating stars. The lower-limits on progenitor ages and the upper-limits on progenitor masses
  are represented by $--$ and $++$ symbols, respectively.}
  \begin{tabular}{@{}llrrrrrr@{}}
  \hline
   WD & Cluster & $W_{\rm{WD}}$ & $t_{\rm{prog}}$ & $M_{{\rm prog}}$ ({\em BaSTI}) & $M_{{\rm prog}}$ (rot.) \label{t:progenitor}\\        
   & &  (M$_{\sun}$) &(Gyr) &  (M$_{\sun}$) & (M$_{\sun}$)  \\
  \hline
VPHAS\,J0804$-$2809 & NGC\,2527     &$0.77_{-0.03}^{+0.03}$&$0.441_{+0.188}^{+0.188}$&$3.06_{-0.35}^{+0.72}$&$3.13_{-0.30}^{+0.70}$\\[1ex]
VPHAS\,J1030$-$5900 & Loden\,143    &$0.65_{-0.02}^{+0.02}$&$0.200_{+0.023}^{+0.023}$&$4.02_{-0.12}^{+0.21}$&$4.22_{-0.18}^{+0.21}$\\[1ex]
VPHAS\,J1103$-$5837 & NGC\,3532     &$1.13_{-0.03}^{+0.03}$&$0.030_{--}^{+0.123}$&$8.80_{-4.31}^{++}$&$9.78_{-5.08}^{++}$\\[1ex]
VPHAS\,J1546$-$5233 & Johansson\,1  &$0.62_{-0.05}^{+0.05}$&$0.437_{+0.196}^{+0.195}$&$3.07_{-0.36}^{+0.75}$&$3.14_{-0.31}^{+0.78}$\\[1ex]
VPHAS\,J1748$-$2914 & Ruprecht\,131 &$0.60_{-0.12}^{+0.16}$&$0.626_{--}^{+0.660}$&$2.72_{-0.64}^{++}$&$2.84_{-0.79}^{++}$\\[1ex]
\hline
\end{tabular}
\end{table*}
\begin{figure}
\includegraphics[width=\linewidth]{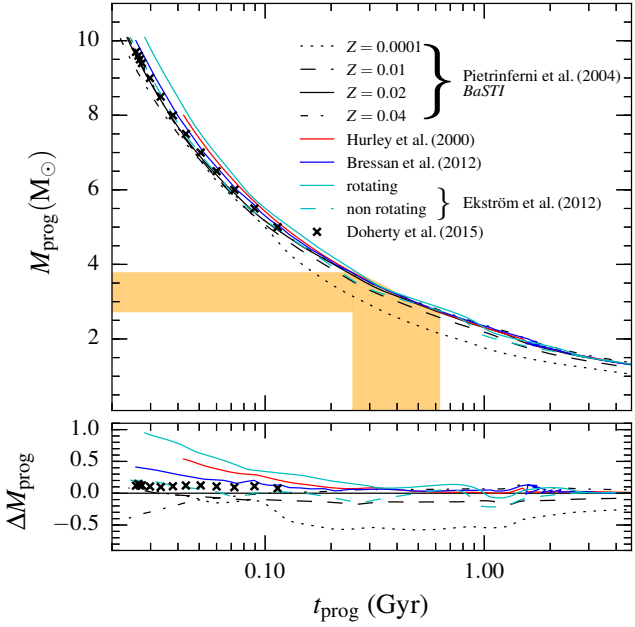}
\caption{Top panel: comparison between different isochrones for stars with $M \leq 10$\,M$_{\sun}$, showing the progenitor mass
as a function of the progenitor lifetime, from the main sequence to the tip of the thermally pulsing AGB. 
The little bump near 2--2.5\,M$_{\sun}$ corresponds to the different evolutionary rate of stars 
experiencing the core helium-flash in degenerate (low-mass range) or non degenerate conditions (high-mass range). 
To illustrate with an example the determination of the progenitor mass, we show the procedure
for VPHAS\,J0804$-$2809 with the light-colour shaded area. Bottom panel: difference between progenitor masses
inferred from different isochrones with respect to the {\em BaSTI} $Z = 0.02$ model, in function of the progenitor lifetime.}
\label{f:progenitor}
\end{figure}

For white dwarfs in open clusters, it is possible to empirically infer the progenitor lifetime, i.e.
the time spent on the main sequence and during the giant phases:
\begin{equation} 
t_{\rm prog} = t_{\rm cluster} - t_{\rm WD}, 
\end{equation}
where $t_{\rm prog}$, $t_{\rm cluster}$, and $t_{\rm WD}$ are the progenitor lifetime, 
cluster age, and white dwarf cooling age, respectively. 

It is possible to estimate the initial mass of the white dwarf progenitor, $M_{i}$, comparing $t_{\rm prog}$
with evolutionary models for single stars. For this purpose, we adopted cluster parameters from the available
literature (Table\,\ref{t:clusters}) and the {\em BaSTI} isochrones \citep[][]{Pietrinferni04}.
The error budget for the progenitor masses takes into account the  uncertainties described in Section\,\ref{chap4} and the cluster age uncertainties.
It is important to note that, at least for progenitor masses below 4\,M$_{\sun}$, and down to 2\,M$_{\sun}$,
different sets of isochrones give similar results. To substantiate this, we give a visual
representation of the progenitor lifetimes for stars of $\leq 10$\,M$_{\sun}$ in Fig.~\ref{f:progenitor},
where we also represent the difference in $M_{i}$ obtain using different sets of isochrones.
To show how the masses are inferred from the models, we display graphically the determination of $M_{i}$ for one of the cluster members (VPHAS\,J0804$-$2809).
The {\em BaSTI} isochrones take into account a standard \citet{Reimers75} parametrisation of mass-loss, with $\eta = 0.4$,
and core convective overshooting during the main sequence, but they do not include other effects like
gravitational settling, radiative acceleration, and rotational mixing. 
The effect of metallicity is relatively subtle, but it becomes evident for very metal poor models ($Z = 0.0001$), 
for which stars less massive than 5\,M$_{\sun}$ evolve much faster.
Although not all the studied clusters have accurate measures of metallicity (Table\,\ref{t:clusters}),
the progenitor age uncertainties are too large to enable a sensible distinction between 
progenitor lifetimes for different metallicities.
Thus, we have added a further term in the error budget, which includes the differences in progenitor lifetimes due to a choice of 
isochrones with  $Z = 0.01,\,0.02,\,0.04$, corresponding to the range of [Fe/H] for $<4$\,Gyr old clusters in the 
Solar neighbourhood \citep[see e.g. Region\,II of table\,2 in][]{Magrini09}. 
In Table\,\ref{t:progenitor}, we list the the progenitor masses of the five cluster members. 

A comparison with the analytical formulation by \citet{Hurley00}, and the PARSEC isochrones \citep{Bressan12},
shows them to favour a slightly slower evolution for stars of $\geq 4$\,M$_{\sun}$. These two models use different efficiencies for the mass-loss
($\eta = 0.5,\,0.2$ respectively), and the PARSEC models include a somewhat more up-to-date physics with a
different Solar model, which determines their Z$_{\sun}$ and mixing-length \citep[see][for a discussion]{Bressan12}.
For comparison in Fig.\,\ref{f:progenitor}, we also plot the lifetimes for the \citet{Doherty15} super-AGB stars and 
the \citet{Ekstrom12} non-rotating models, which all fall in between the \citet{Hurley00} and {\em BaSTI} curves.
A more extreme case, however, is represented by the  \citet{Ekstrom12} rotating  models, which consider an initial
rotation rate on the zero-age main sequence of $0.4$ times the critical escape velocity.
Progenitors of our white dwarfs could have been stars with main sequence masses of $\geq2$\,M$_{\sun}$, i.e. B- or A-type stars.
These are typically fast rotators, whose main sequence lifetime is prolonged in the \citet{Ekstrom12} formulation due to radial mixing of stellar material,
bringing unprocessed hydrogen in to the core. Therefore, for a given mass, a rotating model has a longer-lasting main sequence
than a non-rotating one. The effect of rotation becomes evident for stars with masses larger than 
2\,M$_{\sun}$, and introduces a difference of up to 1\,M$_{\sun}$ when the progenitor mass is estimated at a given progenitor age  (Fig\,\ref{f:progenitor}, bottom panel).
From the point of view of white dwarf structure, stellar rotation is suggested to be important as it could cause a {\em lifting} 
effect that keeps the core temperature of AGB stars below the critical ignition of carbon off-centre, allowing stable, 
massive carbon-oxygen white dwarf to exist \citep[][]{Dominguez96}. 
Since rotation would influence the core mass, and thus the white dwarf structure, 
we also list the progenitor masses interpolated from the \citet{Ekstrom12} rotating models in Table\,\ref{t:progenitor}.

\subsection{Initial-to-final mass relation}
\label{chap5.2}
\begin{figure*}
\includegraphics[width=\linewidth]{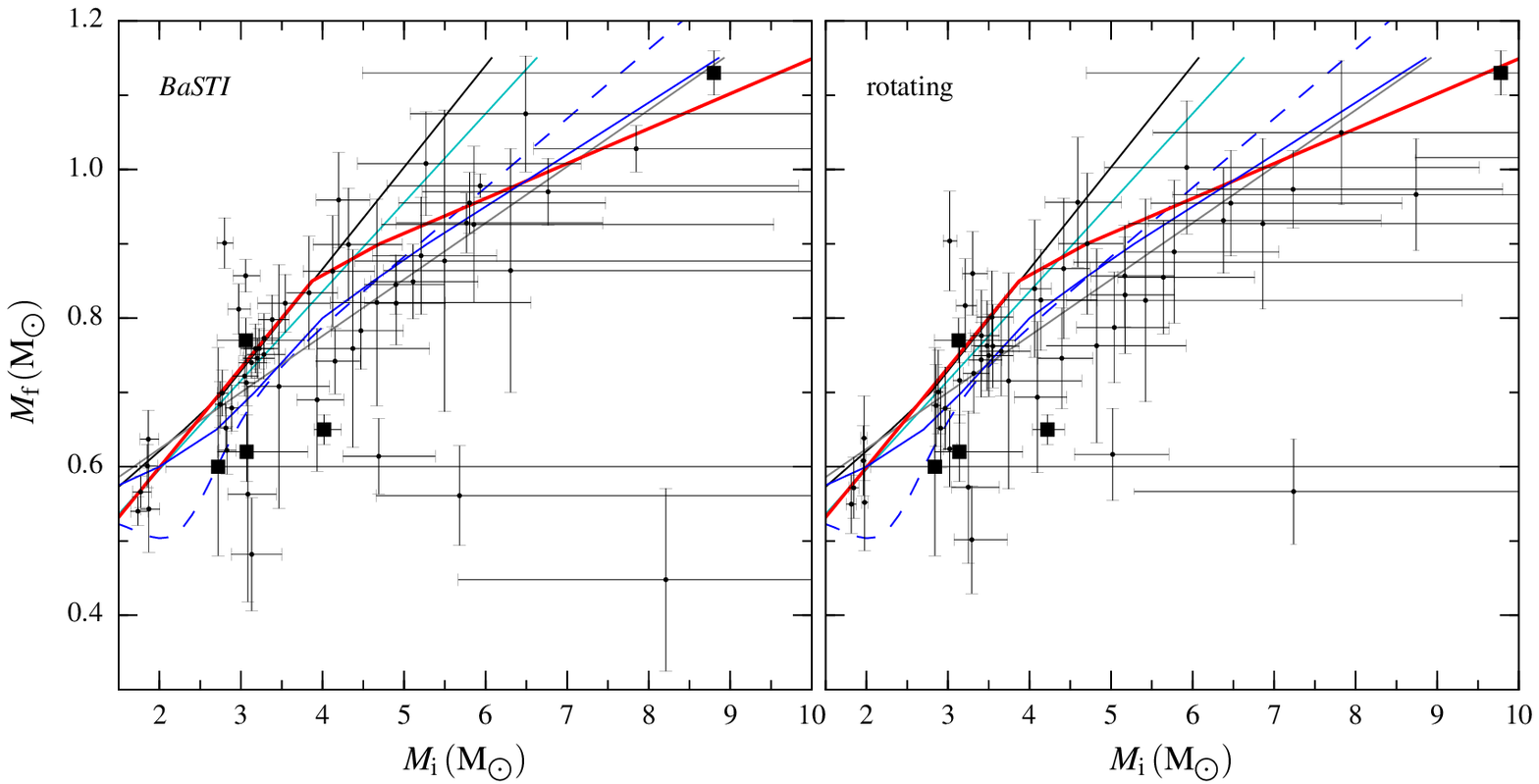}
\caption{Left: initial-to-final mass relation using the data from \citet{Salaris09}, small dots, 
and the five new white dwarfs studied here, large squares. Right: initial-to-final mass relation with progenitor masses computed
from the \citet{Ekstrom12} rotating models. Error bars extending beyond 10\,M$_{\sun}$ indicate upper limits.
Initial-to-final mass relations from the literature are: black, \citet{Catalan08b}; cyan \citet{Casewell09};
red, \citet{Salaris09}; grey, \citet{Gesicki14}; solid-blue, \citet{Weidemann00}.
The dashed-blue curve is the core mass at the first thermal pulse (PARSEC isochrones), 
using the \citet{Marigo13} parametrisation.}
\label{f:ifmr}
\end{figure*}

The initial-to-final mass relation determined from Galactic open clusters suffers from relatively large scatter, 
mostly arising from the mutual interplay of intrinsic stellar properties.
Binary evolution and interactions with other cluster members can also influence stellar evolution,
adding a further source of uncertainty.
All this can be worsened by model-dependent systematics, 
affecting the determination of cluster parameters and stellar evolution.
In Fig.\,\ref{f:ifmr}, we compare the initial and final masses of the five new cluster members 
to the 50 well-established cluster white dwarfs discussed by \citet[][and references therein]{Salaris09}.
The authors used {\em BaSTI} evolutionary models to determine the cluster distances and ages, via interpolation with main sequence isochrones,
and white dwarf cooling ages via \citet[][and references therein]{Salaris10} cooling models.  
Their approach insured that the progenitor ages and masses are estimated from a well defined set of initial and final conditions. 
In the left panel, we plot the initial masses interpolated from {\em BaSTI} isochrones. 
We note satisfying agreement for VPHAS\,J0804$-$2809, VPHAS\,J1103$-$5837,
VPHAS\,J1546$-$5233, with the empirical relations by other authors.

The DAH white dwarf, VPHAS\,J1748$-$2914, overlaps the theoretical curve
representing the core mass at the first thermal pulse \citep[][]{Marigo13},
which is the dominant factor in determining the white dwarf mass \citep[][]{Weidemann00}.
However, for this white dwarf the errors on the progenitor age are too large to derive a
meaningful mass and we only give a lower limit. It is very important to find magnetic white dwarfs in open clusters, as they can be used to  
constrain the mass of their progenitors, yet unknown \citep[][]{Kulebi13},
helping to understand the debated origin of magnetic fields in white dwarfs \citep[][]{Ferrario15}, 

The progenitor mass of the remaining cluster white dwarf,  VPHAS\,J1030$-$5900, 
falls below most of the other white dwarfs and the various initial-to-final mass relation curves.
In the past, a few interpretations have been given to explain such outliers, 
including differential mass-loss on the giant branches due to metallicity,
and binary interactions \citep[][]{Weidemann00}.We note that at least two of the white dwarfs considered in \citet{Salaris09} 
also have large progenitor masses, but final masses below 0.6\,M$_{\sun}$.
As we suggested in Section\,\ref{chap2}, the physical parameters of 
Loden\,143 might be rather uncertain, due to the ambiguous nature of the cluster,
and the initial mass we derive for VPHAS\,J1030$-$5900 may not be correct.
Nevertheless, we would like to stress that binarity may have a relevant effect on the scatter seen in the initial-to-final mass relation, 
especially at the large progenitor-mass end. In fact, white dwarf progenitors of $M_{i} \geq 2$\,M$_{\sun}$ are characterised by 
a relatively high binary fraction in their pre-main sequence \citep[e.g. $\approx 68$--73\,per cent;][]{Baines06}
and later evolutionary stages \citep[25--50\,per cent; see e.g.][]{Abt78, Oudmaijer10}. 
Although we suggested the five new cluster white dwarfs not to have late-type, low-mass companions, 
some other white dwarfs displayed in Fig.\,\ref{f:ifmr} might be or may have been in binary systems.

In the right panel of Fig.\,\ref{f:ifmr}, 
for illustrative purposes we show again the 50 cluster white dwarfs studied in \citet{Salaris09} 
and the five cluster white dwarf from this study, but we determine the progenitor masses of both samples from 
the \citet{Ekstrom12} isochrones. 
As we noted in the previous section, there is a shift towards larger initial masses
(up to 1\,M$_{\sun}$) for stars with $M_{i} \geq 2$\,M$_{\sun}$,
due to the prolonged lifetime as effect of rotational mixing. 
The two panels of Fig.\,\ref{f:ifmr} may not be directly comparable, as the cluster ages that we used
are typically determined from evolutionary models that do not include stellar rotation. 
Given the importance that rotation has for the evolution of the most massive white dwarf progenitors, it should not 
be neglected when studying the evolution of stellar populations, and it would be 
worth to assess in future work its impact on the determination on cluster ages.

\subsubsection{Upper-mass limit of white dwarf progenitors}
\begin{figure}
\includegraphics[width=\linewidth]{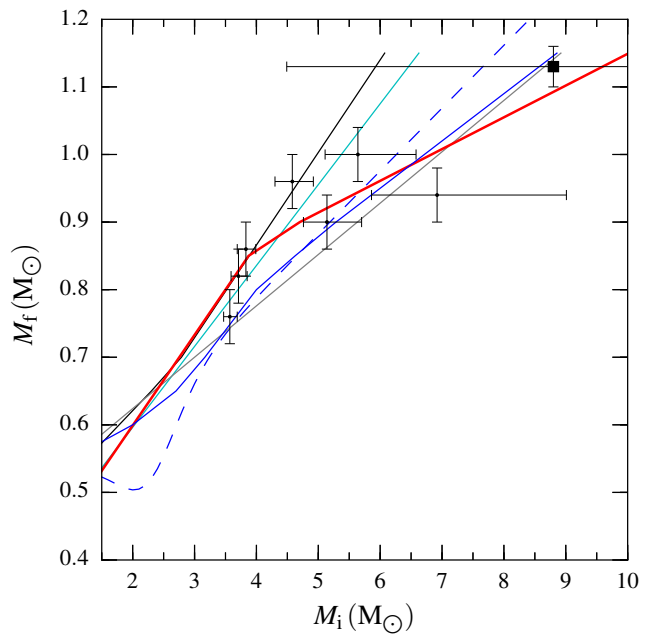}
\caption{Initial-to-final mass relation for the open cluster NGC\,3532 is shown. 
The seven confirmed cluster members  \citep{Dobbie09, Dobbie12} 
and the highest-mass white dwarf, VPHAS\,J1103$-$5837, are plotted. The initial-to-final mass relations
from the literature are depicted as in Fig.\,\ref{f:ifmr}.}
\label{f:d12}
\end{figure}

For the most massive cluster white dwarf in our study, VPHAS\,J1103$-$5837, 
we derived an $8.8_{-4.3}^{+1.2}$\,M$_{\sun}$ progenitor, near the mass-boundary between
white dwarf and neutron star progenitors \citep[][]{Smartt09}. 
To place VPHAS\,J1103$-$5837 in context with the other white dwarfs of NGC\,3532, we display
in Fig.\,\ref{f:d12} their most up-to-date census by \cite{Dobbie09, Dobbie12}. The new white dwarf seems to be genuinely the
cluster member with the most massive progenitor. 
Since the cluster age uncertainty \citep[$\pm 100$\,Myr;][]{Clem11} dominates the error propagation, 
we cannot derive a more accurate measure of the progenitor mass, due to the steep rise of the curves in Fig.\,\ref{f:progenitor}.
However, since massive white dwarfs could also be produced
via binary interaction \citep[][]{Dominguez93} and we do not have information 
on the past history of VPHAS\,J1103$-$5837, we cannot discard that this white dwarf was produced through binary
evolution (merger). 

Considering the single star evolution channel, this result is very interesting, since it adds further empirical evidence 
to previous theoretical and observational works, suggesting the dividing mass to be $M_{i} \gtrsim 7$\,M$_{\sun}$ 
\citep[e.g.][]{GarciaBerro97, Williams09}.
The key ingredients influencing the final mass of white dwarfs are sensitive to stellar parameters like metallicity,
and need to be tested on observed data. In the high-mass range,  theoretical results show that 
high dredge-up efficiency couple to a moderate mass-loss ($\approx 10^{-7}$\,M$_{\sun}$\,yr$^{-1}$),  during the AGB phase,
and appear to dominate the evolution of white dwarf progenitors \citep[e.g.][]{Siess07, Siess10, Doherty15}. 
Observations suggest that  the most intense core-mass growth occurs 
between $M_{i} = 1.6$--3.4\,M$_{\sun}$ (30\,per cent), 
while it is appears to be smaller ($\approx 10$\,per cent) for stars of $M_{i} \approx 4$\,M$_{\sun}$ \citep[][]{Kalirai14}.
However, the core-mass growth of massive white dwarf progenitors still needs to be confirmed.
Thus, the search of other massive cluster white dwarfs should be prioritised, in order to better constrain the high-mass end of the initial-to-final mass relation.
The evolutionary models for super AGB stars ($M_{i} \geq 5$\,M$_{\sun}$) become very resource-consuming, 
due to extensive time- and spatial-resolution requirements for modelling the TP-AGB phase, and some approximations are taken in to account. 
Stars like VPHAS\,J1103$-$5837 could help to constrain the main uncertainties in the models, 
due to the treatment of convection, mass-loss, and third dredge-up efficiency \citep[][and references therein]{Doherty15}.

\section{Summary and conclusions}
\label{chap6}
We proved the efficient selection of white dwarfs from VPHAS+ DR2 $ugr$ photometry,
enabling the study of faint stellar remnants in the most crowded regions of the Galactic plane.
We confirmed 17 white dwarf candidates with VLT/FORS2 spectroscopy.
We identified 13 DA, one DB, two DAH, and one DC white dwarfs. Their atmospheric parameters, masses, ages, and distances,
derived from model atmosphere analysis, suggest that five of them are likely members of open clusters.

The progenitor masses for the five new cluster members are broadly consistent with the known trend of the initial-to-final mass relation.
VPHAS\,J1103$-$5837, in NGC\,3532, is possibly the most massive white
dwarf known in an open cluster ($1.13\pm 0.03$\,M$_{\sun}$), likely with an oxygen-neon core.
Its progenitor mass, $8.8_{-4.3}^{+1.2}$\,M$_{\sun}$,
is close to the mass-divide with core collapse supernovae. 
Finding more massive cluster white dwarfs, like VPHAS\,J1103$-$5837, 
is important to derive firmer constraints at the high progenitor-mass end of the initial-to-final mass relation.
The DAH white dwarf, VPHAS\,J1748$-$2914, is suggested to belong to Ruprecht\,131.
Future observations of this star, with higher S/N, will be needed to confirm its cluster membership and to measure its progenitor mass, now only defined as
a lower limit of 2--3\,M$_{\sun}$.

VPHAS+ and its twin surveys in the Northern hemisphere \citep[IPHAS, and UVEX;][]{Drew05, Groot09} are ideal
tools for the successful identification of the missing population of faint stellar remnants of low- to intermediate-mass stars in the Galactic plane. 
Optical follow-up  spectroscopy,  with moderate resolution and S/N\,$> 20$, is sufficient to confirm the white dwarfs and to measure their atmospheric parameters, 
but higher quality data are necessary if more accurate spectroscopic parallaxes are to be sought. 
The upcoming multi-object spectrographs, WEAVE on WHT \citep[][]{Dalton12}
and 4MOST on VISTA \citep[][]{deJong12}, will play an important role in confirming more cluster white dwarfs and measuring accurate physical parameters.
The ESA {\em Gaia} mission will deliver parallaxes for several hundred thousand 
white dwarfs down to 18--20\,mag \citep[][]{Jordan07, Carrasco14, Gaensicke15}, with an accuracy of $\approx 30$\,per cent \citep[][]{deBruijne15}. 
ESA {\em Gaia} will supply a crucial improvement to open clusters science, as it will determine stellar membership
via the measure of parallaxes and proper motions, allowing the accurate determination of cluster distances and ages, 
and thus significantly improving the study of the initial-to-final mass relation.
\section{Acknowledgements}
Based on data products from observations made with ESO Telescopes at the La
Silla Paranal Observatory under programme ID\,177.D-3023, as part of the VST
Photometric $\Halpha$ Survey of the Southern Galactic plane and Bulge
(\url{www.vphas.eu}), and programme ID\,093.D-0838.

We would like to thank Jorick Vink and Dimitri Veras for useful discussion, 
and Nick Wright for helpful advice on the use of Montage. 
We would like to thank Israel Blanchard, Dimitri Gadotti, 
Patricia Guajardo, and Roger Wesson for their support at Paranal.
We would also like to thank the referee, Pierre Bergeron, for his useful comments. 

The research leading to these results has received funding from the
European Research Council under the European Union's Seventh Framework
Programme (FP/2007-2013) / ERC Grant Agreement n. 320964 (WDTracer).

This research was made possible through the use of the 
AAVSO Photometric All-Sky Survey (APASS), funded by the Robert Martin Ayers Sciences Fund.
The VPHAS+ mosaics were produced with Montage. 
It is funded by the National Science Foundation under Grant Number ACI-1440620, 
and was previously funded by the National Aeronautics and 
Space Administration's Earth Science Technology Office, 
Computation Technologies Project, 
under Cooperative Agreement Number NCC5-626 
between NASA and the California Institute of Technology.

\bibliographystyle{mnras}

\appendix
\section{White dwarf cooling tracks}
\label{app:one}
\begin{table*}
\scriptsize

  \caption{$g$-band absolute magnitudes and colours of DA white dwarfs in the VPHAS+ Vega system.}
\makebox[\textwidth][c]{\begin{tabular}{@{}rrrrrrrrrrrrrrrr@{}}
  \hline
  & \multicolumn{5}{c}{$\log{g}=7.00$}&\multicolumn{5}{c}{$\log{g}=7.50$}&\multicolumn{5}{c}{$\log{g}=8.00$}\\
  $T_{\rm{eff}}$ & $g$ & $(u-g)$ & $(g-r)$ & $(r-i)$ & $(r-\Halpha)$&  $g$ & $(u-g)$ & $(g-r)$ & $(r-i)$ & $(r-\Halpha)$&  $g$ & $(u-g)$ & $(g-r)$ & $(r-i)$ & $(r-\Halpha)$\\ 
  \hline
100000&$6.114$&$-1.603$&$-0.317$&$-0.173$&$0.045$&$7.209$&$-1.603$&$-0.315$&$-0.172$&$0.040$&$8.255$&$-1.604$&$-0.313$&$-0.172$&$0.035$\\
90000&$6.289$&$-1.595$&$-0.313$&$-0.171$&$0.044$&$7.349$&$-1.595$&$-0.311$&$-0.171$&$0.039$&$9.338$&$-1.599$&$-0.307$&$-0.169$&$0.026$\\
80000&$6.473$&$-1.585$&$-0.308$&$-0.169$&$0.042$&$7.511$&$-1.586$&$-0.306$&$-0.169$&$0.036$&$9.450$&$-1.590$&$-0.302$&$-0.167$&$0.022$\\
75000&$6.580$&$-1.579$&$-0.306$&$-0.168$&$0.041$&$7.597$&$-1.580$&$-0.304$&$-0.167$&$0.034$&$9.511$&$-1.585$&$-0.299$&$-0.166$&$0.019$\\
70000&$6.698$&$-1.572$&$-0.303$&$-0.167$&$0.039$&$7.686$&$-1.574$&$-0.301$&$-0.166$&$0.032$&$9.578$&$-1.579$&$-0.295$&$-0.165$&$0.016$\\
65000&$6.818$&$-1.564$&$-0.300$&$-0.165$&$0.037$&$7.781$&$-1.566$&$-0.297$&$-0.165$&$0.030$&$9.649$&$-1.573$&$-0.292$&$-0.163$&$0.013$\\
60000&$6.939$&$-1.555$&$-0.296$&$-0.164$&$0.035$&$7.884$&$-1.557$&$-0.294$&$-0.164$&$0.027$&$9.726$&$-1.565$&$-0.288$&$-0.162$&$0.009$\\
55000&$7.063$&$-1.544$&$-0.292$&$-0.163$&$0.032$&$7.997$&$-1.547$&$-0.289$&$-0.162$&$0.023$&$9.813$&$-1.555$&$-0.283$&$-0.160$&$0.003$\\
50000&$7.197$&$-1.530$&$-0.287$&$-0.161$&$0.028$&$8.127$&$-1.533$&$-0.284$&$-0.160$&$0.018$&$9.911$&$-1.543$&$-0.277$&$-0.158$&$-0.003$\\
45000&$7.352$&$-1.512$&$-0.281$&$-0.158$&$0.022$&$8.274$&$-1.516$&$-0.278$&$-0.158$&$0.012$&$10.027$&$-1.528$&$-0.269$&$-0.155$&$-0.013$\\
40000&$7.540$&$-1.485$&$-0.273$&$-0.155$&$0.014$&$8.451$&$-1.490$&$-0.269$&$-0.154$&$0.002$&$10.172$&$-1.505$&$-0.259$&$-0.151$&$-0.026$\\
35000&$7.803$&$-1.441$&$-0.261$&$-0.150$&$0.001$&$8.685$&$-1.448$&$-0.256$&$-0.149$&$-0.014$&$10.372$&$-1.469$&$-0.244$&$-0.146$&$-0.047$\\
30000&$8.182$&$-1.347$&$-0.240$&$-0.141$&$-0.024$&$9.058$&$-1.360$&$-0.233$&$-0.140$&$-0.043$&$10.695$&$-1.393$&$-0.216$&$-0.135$&$-0.087$\\
28000&$8.365$&$-1.293$&$-0.227$&$-0.134$&$-0.033$&$9.235$&$-1.307$&$-0.219$&$-0.133$&$-0.054$&$10.858$&$-1.346$&$-0.200$&$-0.128$&$-0.104$\\
26000&$8.549$&$-1.236$&$-0.210$&$-0.126$&$-0.038$&$9.413$&$-1.251$&$-0.202$&$-0.125$&$-0.062$&$11.025$&$-1.294$&$-0.180$&$-0.120$&$-0.115$\\
24000&$8.734$&$-1.174$&$-0.192$&$-0.117$&$-0.043$&$9.591$&$-1.192$&$-0.183$&$-0.116$&$-0.068$&$11.186$&$-1.238$&$-0.159$&$-0.111$&$-0.123$\\
22000&$8.928$&$-1.105$&$-0.172$&$-0.108$&$-0.049$&$9.776$&$-1.124$&$-0.162$&$-0.106$&$-0.075$&$11.358$&$-1.176$&$-0.135$&$-0.101$&$-0.133$\\
20000&$9.134$&$-1.023$&$-0.150$&$-0.098$&$-0.058$&$9.973$&$-1.045$&$-0.139$&$-0.096$&$-0.085$&$11.542$&$-1.103$&$-0.108$&$-0.090$&$-0.145$\\
19000&$9.246$&$-0.977$&$-0.138$&$-0.092$&$-0.063$&$10.078$&$-0.999$&$-0.126$&$-0.090$&$-0.091$&$11.640$&$-1.062$&$-0.093$&$-0.083$&$-0.153$\\
18000&$9.364$&$-0.925$&$-0.125$&$-0.086$&$-0.069$&$10.189$&$-0.949$&$-0.112$&$-0.084$&$-0.099$&$11.744$&$-1.017$&$-0.077$&$-0.076$&$-0.163$\\
17000&$9.487$&$-0.867$&$-0.111$&$-0.079$&$-0.077$&$10.306$&$-0.894$&$-0.096$&$-0.077$&$-0.108$&$11.856$&$-0.969$&$-0.058$&$-0.068$&$-0.174$\\
16000&$9.621$&$-0.803$&$-0.095$&$-0.072$&$-0.086$&$10.432$&$-0.833$&$-0.079$&$-0.069$&$-0.119$&$11.975$&$-0.922$&$-0.035$&$-0.059$&$-0.183$\\
15000&$9.766$&$-0.732$&$-0.077$&$-0.063$&$-0.097$&$10.567$&$-0.767$&$-0.058$&$-0.060$&$-0.132$&$12.105$&$-0.883$&$-0.008$&$-0.046$&$-0.187$\\
14000&$9.924$&$-0.654$&$-0.055$&$-0.053$&$-0.111$&$10.716$&$-0.698$&$-0.033$&$-0.049$&$-0.146$&$12.239$&$-0.857$&$0.023$&$-0.029$&$-0.188$\\
13000&$10.101$&$-0.573$&$-0.027$&$-0.041$&$-0.126$&$10.884$&$-0.634$&$-0.001$&$-0.034$&$-0.156$&$12.367$&$-0.837$&$0.053$&$-0.012$&$-0.188$\\
12000&$10.303$&$-0.500$&$0.011$&$-0.024$&$-0.137$&$11.065$&$-0.602$&$0.044$&$-0.013$&$-0.158$&$12.547$&$-0.810$&$0.093$&$0.010$&$-0.180$\\
11000&$10.524$&$-0.481$&$0.068$&$0.003$&$-0.134$&$11.273$&$-0.595$&$0.096$&$0.015$&$-0.146$&$12.793$&$-0.780$&$0.138$&$0.039$&$-0.157$\\
10000&$10.839$&$-0.495$&$0.141$&$0.043$&$-0.101$&$11.596$&$-0.602$&$0.163$&$0.057$&$-0.102$&$13.123$&$-0.774$&$0.194$&$0.077$&$-0.108$\\
9500&$11.053$&$-0.518$&$0.183$&$0.070$&$-0.070$&$11.802$&$-0.620$&$0.203$&$0.082$&$-0.069$&$13.320$&$-0.774$&$0.226$&$0.100$&$-0.075$\\
9000&$11.297$&$-0.552$&$0.230$&$0.099$&$-0.031$&$12.031$&$-0.641$&$0.246$&$0.108$&$-0.031$&$13.530$&$-0.770$&$0.262$&$0.122$&$-0.042$\\
8500&$11.565$&$-0.583$&$0.279$&$0.129$&$0.009$&$12.277$&$-0.654$&$0.291$&$0.135$&$0.007$&$13.756$&$-0.756$&$0.301$&$0.145$&$-0.012$\\
8000&$11.852$&$-0.600$&$0.333$&$0.159$&$0.048$&$12.541$&$-0.652$&$0.341$&$0.163$&$0.041$&$13.996$&$-0.727$&$0.346$&$0.169$&$0.017$\\
7500&$12.159$&$-0.594$&$0.392$&$0.191$&$0.081$&$12.824$&$-0.629$&$0.397$&$0.193$&$0.069$&$14.257$&$-0.678$&$0.399$&$0.198$&$0.043$\\
7000&$12.492$&$-0.559$&$0.460$&$0.225$&$0.105$&$13.135$&$-0.580$&$0.462$&$0.227$&$0.096$&$14.544$&$-0.595$&$0.463$&$0.229$&$0.074$\\
6500&$12.861$&$-0.485$&$0.537$&$0.264$&$0.134$&$13.482$&$-0.489$&$0.541$&$0.265$&$0.126$&$14.877$&$-0.453$&$0.547$&$0.266$&$0.113$\\
6000&$13.286$&$-0.352$&$0.637$&$0.310$&$0.166$&$13.888$&$-0.325$&$0.644$&$0.311$&$0.162$&$15.274$&$-0.187$&$0.654$&$0.310$&$0.156$\\
\vspace{0.1cm}\\
  & \multicolumn{5}{c}{$\log{g}=8.50$}&\multicolumn{5}{c}{$\log{g}=9.00$}&&&&&\\
  $T_{\rm{eff}}$ &  $g$ & $(u-g)$ & $(g-r)$ & $(r-i)$ & $(r-\Halpha)$& $g$ & $(u-g)$ & $(g-r)$ & $(r-i)$ & $(r-\Halpha)$&&&&&\\
\vspace{0.1cm}\\
100000&$9.238$&$-1.606$&$-0.311$&$-0.171$&$0.029$&$10.299$&$-1.608$&$-0.309$&$-0.170$&$0.022$&&&&&\\
90000&$9.338$&$-1.599$&$-0.307$&$-0.169$&$0.026$&$10.388$&$-1.601$&$-0.305$&$-0.168$&$0.018$\\
80000&$9.450$&$-1.590$&$-0.302$&$-0.167$&$0.022$&$10.492$&$-1.593$&$-0.299$&$-0.166$&$0.014$\\
75000&$9.511$&$-1.585$&$-0.299$&$-0.166$&$0.019$&$10.549$&$-1.589$&$-0.296$&$-0.165$&$0.011$\\
70000&$9.578$&$-1.579$&$-0.295$&$-0.165$&$0.016$&$10.610$&$-1.583$&$-0.293$&$-0.163$&$0.008$\\
65000&$9.649$&$-1.573$&$-0.292$&$-0.163$&$0.013$&$10.677$&$-1.577$&$-0.289$&$-0.162$&$0.004$\\
60000&$9.726$&$-1.565$&$-0.288$&$-0.162$&$0.009$&$10.752$&$-1.570$&$-0.284$&$-0.160$&$-0.001$\\
55000&$9.813$&$-1.555$&$-0.283$&$-0.160$&$0.003$&$10.834$&$-1.561$&$-0.279$&$-0.158$&$-0.007$\\
50000&$9.911$&$-1.543$&$-0.277$&$-0.158$&$-0.003$&$10.928$&$-1.550$&$-0.273$&$-0.156$&$-0.015$\\
45000&$10.027$&$-1.528$&$-0.269$&$-0.155$&$-0.013$&$11.039$&$-1.535$&$-0.265$&$-0.153$&$-0.026$\\
40000&$10.172$&$-1.505$&$-0.259$&$-0.151$&$-0.026$&$11.178$&$-1.515$&$-0.254$&$-0.149$&$-0.041$\\
35000&$10.372$&$-1.469$&$-0.244$&$-0.146$&$-0.047$&$11.372$&$-1.481$&$-0.237$&$-0.143$&$-0.065$\\
30000&$10.695$&$-1.393$&$-0.216$&$-0.135$&$-0.087$&$11.677$&$-1.413$&$-0.207$&$-0.132$&$-0.110$\\
28000&$10.858$&$-1.346$&$-0.200$&$-0.128$&$-0.104$&$11.838$&$-1.370$&$-0.189$&$-0.125$&$-0.130$\\
26000&$11.025$&$-1.294$&$-0.180$&$-0.120$&$-0.115$&$12.005$&$-1.320$&$-0.167$&$-0.116$&$-0.143$\\
24000&$11.186$&$-1.238$&$-0.159$&$-0.111$&$-0.123$&$12.163$&$-1.267$&$-0.145$&$-0.108$&$-0.153$\\
22000&$11.358$&$-1.176$&$-0.135$&$-0.101$&$-0.133$&$12.335$&$-1.208$&$-0.119$&$-0.097$&$-0.163$\\
20000&$11.542$&$-1.103$&$-0.108$&$-0.090$&$-0.145$&$12.519$&$-1.139$&$-0.091$&$-0.085$&$-0.176$\\
19000&$11.640$&$-1.062$&$-0.093$&$-0.083$&$-0.153$&$12.618$&$-1.101$&$-0.074$&$-0.078$&$-0.185$\\
18000&$11.744$&$-1.017$&$-0.077$&$-0.076$&$-0.163$&$12.723$&$-1.061$&$-0.056$&$-0.071$&$-0.194$\\
17000&$11.856$&$-0.969$&$-0.058$&$-0.068$&$-0.174$&$12.835$&$-1.023$&$-0.034$&$-0.061$&$-0.201$\\
16000&$11.975$&$-0.922$&$-0.035$&$-0.059$&$-0.183$&$12.950$&$-0.993$&$-0.010$&$-0.048$&$-0.203$\\
15000&$12.105$&$-0.883$&$-0.008$&$-0.046$&$-0.187$&$13.065$&$-0.971$&$0.016$&$-0.034$&$-0.203$\\
14000&$12.239$&$-0.857$&$0.023$&$-0.029$&$-0.188$&$13.175$&$-0.950$&$0.040$&$-0.021$&$-0.203$\\
13000&$12.367$&$-0.837$&$0.053$&$-0.012$&$-0.188$&$13.323$&$-0.922$&$0.070$&$-0.002$&$-0.199$\\
12000&$12.547$&$-0.810$&$0.093$&$0.010$&$-0.180$&$13.524$&$-0.885$&$0.109$&$0.021$&$-0.190$\\
11000&$12.793$&$-0.780$&$0.138$&$0.039$&$-0.157$&$13.780$&$-0.853$&$0.150$&$0.050$&$-0.165$\\
10000&$13.123$&$-0.774$&$0.194$&$0.077$&$-0.108$&$14.112$&$-0.841$&$0.201$&$0.086$&$-0.115$\\
9500&$13.320$&$-0.774$&$0.226$&$0.100$&$-0.075$&$14.303$&$-0.833$&$0.231$&$0.106$&$-0.085$\\
9000&$13.530$&$-0.770$&$0.262$&$0.122$&$-0.042$&$14.505$&$-0.818$&$0.264$&$0.126$&$-0.056$\\
8500&$13.756$&$-0.756$&$0.301$&$0.145$&$-0.012$&$14.719$&$-0.793$&$0.301$&$0.148$&$-0.027$\\
8000&$13.996$&$-0.727$&$0.346$&$0.169$&$0.017$&$14.950$&$-0.752$&$0.345$&$0.172$&$-0.002$\\
7500&$14.257$&$-0.678$&$0.399$&$0.198$&$0.043$&$15.199$&$-0.687$&$0.396$&$0.199$&$0.028$\\
7000&$14.544$&$-0.595$&$0.463$&$0.229$&$0.074$&$15.481$&$-0.582$&$0.463$&$0.229$&$0.064$\\
6500&$14.877$&$-0.453$&$0.547$&$0.266$&$0.113$&$15.814$&$-0.404$&$0.550$&$0.266$&$0.109$\\
6000&$15.274$&$-0.187$&$0.654$&$0.310$&$0.156$&$16.211$&$-0.043$&$0.665$&$0.308$&$0.156$\\
  \hline
\end{tabular}}
\end{table*}

\begin{table*}
\scriptsize
  \caption{$g$-band absolute magnitudes and colours of DB white dwarfs in the VPHAS+ Vega system.}
\makebox[\textwidth][c]{  \begin{tabular}{@{}rrrrrrrrrrrrrrrr@{}}
   \hline
  & \multicolumn{5}{c}{$\log{g}=7.00$}&\multicolumn{5}{c}{$\log{g}=7.50$}&\multicolumn{5}{c}{$\log{g}=8.00$}\\
  $T_{\rm{eff}}$ & $g$ & $(u-g)$ & $(g-r)$ & $(r-i)$ & $(r-\Halpha)$& $g$ & $(u-g)$ & $(g-r)$ & $(r-i)$ & $(r-\Halpha)$& $g$ & $(u-g)$ & $(g-r)$ & $(r-i)$ & $(r-\Halpha)$\\ 
  \hline
40000&$7.672$&$-1.467$&$-0.260$&$-0.148$&$0.073$&$8.710$&$-1.466$&$-0.254$&$-0.144$&$0.071$&$9.557$&$-1.469$&$-0.250$&$-0.141$&$0.069$\\
35000&$8.016$&$-1.407$&$-0.234$&$-0.135$&$0.076$&$9.008$&$-1.409$&$-0.228$&$-0.132$&$0.074$&$9.824$&$-1.413$&$-0.223$&$-0.128$&$0.070$\\
30000&$8.375$&$-1.334$&$-0.205$&$-0.122$&$0.078$&$9.317$&$-1.337$&$-0.198$&$-0.118$&$0.075$&$10.105$&$-1.345$&$-0.192$&$-0.112$&$0.070$\\
28000&$8.530$&$-1.299$&$-0.191$&$-0.115$&$0.079$&$9.447$&$-1.304$&$-0.183$&$-0.111$&$0.075$&$10.225$&$-1.317$&$-0.177$&$-0.105$&$0.070$\\
26000&$8.691$&$-1.263$&$-0.175$&$-0.108$&$0.079$&$9.585$&$-1.273$&$-0.168$&$-0.103$&$0.075$&$10.349$&$-1.291$&$-0.161$&$-0.096$&$0.071$\\
24000&$8.861$&$-1.227$&$-0.158$&$-0.099$&$0.080$&$9.730$&$-1.242$&$-0.150$&$-0.094$&$0.076$&$10.467$&$-1.269$&$-0.145$&$-0.088$&$0.073$\\
22000&$9.043$&$-1.191$&$-0.139$&$-0.090$&$0.082$&$9.863$&$-1.218$&$-0.134$&$-0.087$&$0.078$&$10.553$&$-1.257$&$-0.133$&$-0.083$&$0.075$\\
20000&$9.179$&$-1.174$&$-0.127$&$-0.087$&$0.082$&$9.957$&$-1.212$&$-0.123$&$-0.083$&$0.080$&$10.662$&$-1.248$&$-0.117$&$-0.077$&$0.080$\\
19000&$9.245$&$-1.174$&$-0.119$&$-0.085$&$0.084$&$10.039$&$-1.209$&$-0.112$&$-0.078$&$0.085$&$10.753$&$-1.242$&$-0.105$&$-0.072$&$0.085$\\
18000&$9.351$&$-1.174$&$-0.106$&$-0.079$&$0.090$&$10.149$&$-1.207$&$-0.097$&$-0.072$&$0.091$&$10.862$&$-1.236$&$-0.090$&$-0.067$&$0.092$\\
17000&$9.466$&$-1.177$&$-0.088$&$-0.070$&$0.096$&$10.279$&$-1.205$&$-0.080$&$-0.064$&$0.098$&$10.989$&$-1.230$&$-0.073$&$-0.060$&$0.100$\\
16000&$9.626$&$-1.178$&$-0.067$&$-0.060$&$0.104$&$10.427$&$-1.201$&$-0.059$&$-0.055$&$0.106$&$11.133$&$-1.221$&$-0.054$&$-0.052$&$0.108$\\
15000&$9.805$&$-1.173$&$-0.040$&$-0.048$&$0.111$&$10.595$&$-1.191$&$-0.034$&$-0.044$&$0.113$&$11.295$&$-1.207$&$-0.031$&$-0.041$&$0.114$\\
14000&$10.006$&$-1.159$&$-0.009$&$-0.032$&$0.118$&$10.779$&$-1.172$&$-0.004$&$-0.028$&$0.119$&$11.471$&$-1.184$&$-0.003$&$-0.027$&$0.120$\\
13000&$10.225$&$-1.134$&$0.029$&$-0.013$&$0.125$&$10.982$&$-1.142$&$0.031$&$-0.011$&$0.125$&$11.665$&$-1.149$&$0.032$&$-0.010$&$0.126$\\
12000&$10.466$&$-1.095$&$0.071$&$0.009$&$0.132$&$11.203$&$-1.099$&$0.072$&$0.010$&$0.132$&$11.877$&$-1.102$&$0.072$&$0.011$&$0.132$\\
11000&$10.732$&$-1.033$&$0.118$&$0.032$&$0.139$&$11.449$&$-1.034$&$0.119$&$0.032$&$0.138$&$12.115$&$-1.035$&$0.119$&$0.032$&$0.138$\\
10000&$11.048$&$-0.950$&$0.180$&$0.061$&$0.148$&$11.745$&$-0.950$&$0.180$&$0.061$&$0.147$&$12.403$&$-0.950$&$0.180$&$0.060$&$0.148$\\

\vspace{0.1cm}\\
  & \multicolumn{5}{c}{$\log{g}=8.50$}&\multicolumn{5}{c}{$\log{g}=9.00$}&&&&&\\
  $T_{\rm{eff}}$ &  $g$ & $(u-g)$ & $(g-r)$ & $(r-i)$ & $(r-\Halpha)$& $g$ & $(u-g)$ & $(g-r)$ & $(r-i)$ & $(r-\Halpha)$&&&&&\\
\vspace{0.1cm}\\
40000&$10.403$&$-1.474$&$-0.246$&$-0.136$&$0.066$&$11.396$&$-1.479$&$-0.245$&$-0.130$&$0.062$\\
35000&$10.661$&$-1.420$&$-0.219$&$-0.122$&$0.066$&$11.644$&$-1.429$&$-0.218$&$-0.115$&$0.061$\\
30000&$10.933$&$-1.359$&$-0.188$&$-0.105$&$0.065$&$11.898$&$-1.378$&$-0.187$&$-0.097$&$0.062$\\
28000&$11.042$&$-1.336$&$-0.173$&$-0.097$&$0.066$&$11.989$&$-1.362$&$-0.173$&$-0.090$&$0.065$\\
26000&$11.145$&$-1.317$&$-0.157$&$-0.089$&$0.069$&$12.067$&$-1.348$&$-0.161$&$-0.083$&$0.068$\\
24000&$11.232$&$-1.303$&$-0.144$&$-0.083$&$0.071$&$12.135$&$-1.336$&$-0.152$&$-0.078$&$0.070$\\
22000&$11.306$&$-1.294$&$-0.134$&$-0.078$&$0.074$&$12.240$&$-1.320$&$-0.137$&$-0.071$&$0.075$\\
20000&$11.448$&$-1.275$&$-0.114$&$-0.071$&$0.081$&$12.407$&$-1.301$&$-0.115$&$-0.065$&$0.084$\\
19000&$11.545$&$-1.266$&$-0.103$&$-0.067$&$0.086$&$12.511$&$-1.291$&$-0.102$&$-0.063$&$0.090$\\
18000&$11.657$&$-1.257$&$-0.090$&$-0.063$&$0.094$&$12.629$&$-1.281$&$-0.089$&$-0.060$&$0.097$\\
17000&$11.784$&$-1.250$&$-0.075$&$-0.058$&$0.101$&$12.760$&$-1.269$&$-0.074$&$-0.056$&$0.104$\\
16000&$11.927$&$-1.241$&$-0.058$&$-0.051$&$0.108$&$12.903$&$-1.252$&$-0.055$&$-0.049$&$0.110$\\
15000&$12.084$&$-1.222$&$-0.035$&$-0.041$&$0.114$&$13.058$&$-1.229$&$-0.032$&$-0.039$&$0.116$\\
14000&$12.259$&$-1.195$&$-0.005$&$-0.027$&$0.120$&$13.226$&$-1.197$&$-0.003$&$-0.025$&$0.121$\\
13000&$12.445$&$-1.159$&$0.024$&$-0.012$&$0.125$&$13.409$&$-1.156$&$0.032$&$-0.008$&$0.126$\\
12000&$12.662$&$-1.111$&$0.063$&$0.003$&$0.130$&$13.614$&$-1.104$&$0.072$&$0.011$&$0.132$\\
11000&$12.892$&$-1.035$&$0.119$&$0.032$&$0.139$&$13.848$&$-1.035$&$0.120$&$0.032$&$0.139$\\
10000&$13.176$&$-0.950$&$0.181$&$0.059$&$0.148$&$14.130$&$-0.950$&$0.181$&$0.059$&$0.148$\\
\hline
\end{tabular}}
\end{table*}

\bsp	
\label{lastpage}
\end{document}